\documentclass[aps,pra,reprint,showkeys,nofootinbib,groupedaddress,floatfix]{revtex4-2}
\usepackage{amsmath,amssymb,amsthm}
\usepackage{bm}
\usepackage[sans]{dsfont}
\usepackage{times}
\usepackage{mathtools}
\usepackage[all]{xy}
\theoremstyle{plain}
\newtheorem{theorem}{Theorem}
\newtheorem{proposition}[theorem]{Proposition}

\theoremstyle{remark}
\newtheorem{remark}{Remark}
\newtheorem{assumption}{Assumption}

\theoremstyle{definition}

\newcommand{\Var}{\operatorname{Var}}
\newcommand{\Cov}{\operatorname{Cov}}

\newcommand{\Tr}{\operatorname{Tr}}
\newcommand{\rmd}{\mathrm{d}}
\newcommand{\rme}{\mathrm{e}}
\newcommand{\rmi}{\mathrm{i}}
\newcommand{\RE}{\mathrm{\,Re\,}}
\newcommand{\IM}{\mathrm{\,Im\,}}
\newcommand{\Ebb}{\mathbb{E}}
\newcommand{\Rbb}{\mathbb{R}}
\newcommand{\Cbb}{\mathbb{C}}
\newcommand{\om}{{\omega_{\rm m}}}

\newcommand{\omq}{{\omega_{\rm m}^{\,2}}}
\newcommand{\Om}{{\Omega_{\rm m}}}
\newcommand{\Omq}{{\Omega_{\rm m}^{\,2}}}
\newcommand{\mgamq}{{\gamma_{\rm m}^{\,2}}}
\newcommand{\mgam}{{\gamma_{\rm m}}}
\newcommand{\rmm}{\mathrm{m}}
\newcommand{\ind}{\mathtt{1}}

\newcommand{\norm}[1]{\left\Vert#1\right\Vert}
\newcommand{\abs}[1]{\left\vert#1\right\vert}

\newcommand{\Hcal}{\mathcal{H}}

\newcommand{\Ucal}{\mathcal{U}}

 \newcommand{\Wcal}{\mathcal{W}}

\begin{document}
\title{Quantum optomechanical system in a Mach-Zehnder interferometer}
\author{Alberto Barchielli}
\altaffiliation{Istituto Nazionale di Fisica Nucleare (INFN), Sezione di Milano}
\altaffiliation{Istituto Nazionale di Alta Matematica (INDAM-GNAMPA)}
\email{alberto.barchielli@polimi.it}
\affiliation{Politecnico di Milano, Dipartimento di Matematica, piazza Leonardo da Vinci 32, 20133 Milano, Italy}

\author{Matteo Gregoratti}
\email{matteo.gregoratti@polimi.it }
\affiliation{Politecnico di Milano, Dipartimento di Matematica,
piazza Leonardo da Vinci 32, 20133 Milano, Italy}

\date{\today}

\begin{abstract} We consider an oscillating micromirror replacing one of the two fixed mirrors of a Mach-Zehnder interferometer. In this ideal optical set-up the quantum oscillator is subjected to the radiation pressure interaction of travelling light waves, no cavity is involved. The aim of  this configuration is to show that squeezed light can be generated by pure scattering on a quantum system, without involving a cavity. The squeezing  can be detected at the output ports of the interferometer either by direct detection or by measuring the spectrum of the difference current.
We use the quantum-stochastic Schr\"odinger equation (Hudson-Parthasarathy equation) to model the global evolution. Indeed, it can describe the scattering of photons and the resulting radiation pressure interaction on the quantum oscillator. Moreover, it allows to consider also the interaction with a thermal bath, so that it can describe also the damping of the harmonic oscillator and non-Markovian thermal effects.
In this way we have a unitary dynamics giving the evolution of oscillator and fields. The Bose fields of quantum stochastic calculus and the related generalized Weyl operators allow to describe the whole optical circuit. By working in the Heisenberg picture, the quantum Langevin equations for position and momentum and the output fields arise, which are used to describe the monitoring in continuous time of the light at the output ports. In the case of strong laser and weak radiation pressure interaction  highly non-classical light is produced, and this  can be revealed either by direct detection (a negative Mandel $Q$-parameter is found), either by the intensity spectrum of the difference current of two photodetector; in the second case a nearly complete cancellation of the shot noise can be reached. In this last case it appears that the Mach-Zehnder configuration together with the detection of the difference current corresponds to an homodyne detection scheme, so that we can say that the apparatus is measuring the ``spectrum of squeezing''.
\end{abstract}


\maketitle

\tableofcontents

\section{Introduction}\label{intro}
Quantum optomechanical systems, such as oscillating micromirrors interacting with the light by radiation pressure, have been a very active field of theoretical and experimental research. It is usual to consider such a kind of systems coupled with discrete modes of light in a cavity \cite{JacTWS99,GioV01,GMVT09,SN-P13,Chen13,asp14,GroTru15,BarV15,BM16,Vit+18,ZKRDV18,Vitali12}; indeed, the term of cavity optomechanics is often used. In principle, also the case of travelling waves reflected by a micromirror can be considered \cite{Bar16}; some of the effects typical of cavity optomechanics can be found also in this case.

Here we want to show that, ideally, the pure reflection of the light on an oscillating quantum micromirror can generate squeezed light out of ``classical'' coherent light. As discussed in \cite[Sect.\ 4.5]{BM16} the ponderomotive interaction produces intensity-dependent phase shifts which give rise to optical squeezing. Here we show that this is possible also without the presence of a cavity; the pure scattering of the light on the quantum oscillator can transform the coherent input light in strongly squeezed output light.

Our interest here is not directly in the motion of the quantum oscillator, but in the properties of the output light, in particular in detecting its squeezing.  As we expect the interaction with the quantum system to affect mainly the phase of the light, we insert the quantum micromirror in the place of one of the mirrors of a Mach-Zehnder interferometer (MZI) \cite{HR06,WisM10}; in this way we realize a phase sesitive optical circuit, which provides the interference of the beam of interest with a reference beam, in a way very similar to  an homodyne detection scheme. At the output ports of the MZI, we can think to use direct detection and to check for  \emph{sub-Poissonian statistics} \cite{M79,Car99}; or we can relay on homodyne detection to get the \emph{spectrum of squeezing} \cite[Sect.\ 9.3]{Car08}.

As in \cite{Bar16}, we use quantum stochastic calculus (QSC) and the Hudson-Parthasarathy (HP) equation \cite{HudP84,Parthas92,Bar06} to model the unitary evolution of quantum oscillator and light field. This equation can describe absorbtion/emission of quanta from a quantum system (the oscillator in our case), and we use this feature to introduce the interaction with a thermal bath, giving damping and heating. Moreover, this equation allows also to describe scattering of field quanta \cite{BarP02,BarGQP13,GJN12,JG15} and it is this feature which is used to model the pure reflection of light. By using QSC and HP-equation we can relay on the associated quantum Langevin equations to study the motion of the oscillator (without the need of some additional Markov approximation). On the other side, the related notion of \emph{output fields} \cite{Bar06,GarC85,Bar86,ZolG97} allows to describe the effects of the quantum micromirror on the reflected light.

QSC involves continuous Bose fields, by which we can also model travelling waves of quantum optical fields \cite{ZGVit15}. Moreover, by the related generalized Weyl operators \cite{Parthas92,Bar06}, it is possible to describe the action of linear optical elements on the light and to construct optical circuits \cite{ASth}, as the MZI in the present work.
Finally,  by introducing suitable compatible field observables, we can describe the monitoring of the output light in continuous time (direct, heterodyne, homodyne detection)  \cite{Bar06,GarC85,Bar86,ZolG97,GarZ00,Car99,Car08,BarG08b,BarG12,BarG13}.

\begin{figure*}
\begin{center}
\[
\xymatrix{
*+[F]{\text{\begin{scriptsize}laser\end{scriptsize}}}\ar@{~>}[drr]^{A_1(t)}& & &\ar@{=}[rr] &\ar@{~>}[drr]^{C_2(t)} & {\text{\begin{scriptsize}PS\end{scriptsize}}}                 &&&*+[F]{\text{\begin{scriptsize}photo-counter 1\end{scriptsize}}}\ar@{->}[d]^{I_1(t)}
\\
& {\text{\begin{scriptsize}BS1\end{scriptsize}}}\ar@{-}[rr]
&\ar@{~>}[urr]^{B_2(t)} \ar@{~>}[drr]_{B_1(t)}& & & \ar@{-}[rr]&\ar@{~>}[urr]^{D_1(t)}\ar@{~>}[drr]_{D_2(t)}&{\text{\begin{scriptsize}BS2\end{scriptsize}}}&
*o+[F]{\text{\begin{scriptsize}$
\begin{matrix}\text{post}\\ \text{processing}\end{matrix}$\end{scriptsize}}}
\\
\text{\begin{scriptsize}vacuum\end{scriptsize}}\ar@{.>}[urr]_{A_2(t)}& & &\text{\begin{scriptsize}\phantom{QO}\end{scriptsize}} \ar@{o-o}[rr] &
\ar@{~>}[urr]_{C_1(t)} &{\text{\begin{scriptsize}QO\end{scriptsize}}}
&&&*+[F]{\text{\begin{scriptsize}photo-counter 2\end{scriptsize}}}\ar@{->}[u]_{I_2(t)}
\\
&&& & *+[F]{\text{\begin{scriptsize}thermal  bath\end{scriptsize}} }\ar@2{<~>}[u]_{\text{\begin{scriptsize}phonons\end{scriptsize}}}^{A_3(t)}
&&&&}
\]
\end{center}
\caption{The optical circuit. BS: beam splitter. PS: phase shifter and fixed mirror. QO: reflecting quantum oscillator. $A_j(t)$, $B_j(t)$, $C_j(t)$, $D_j(t)$: bosonic quantum fields. $I_j(t)$: output currents.} \label{MZI}
\end{figure*}
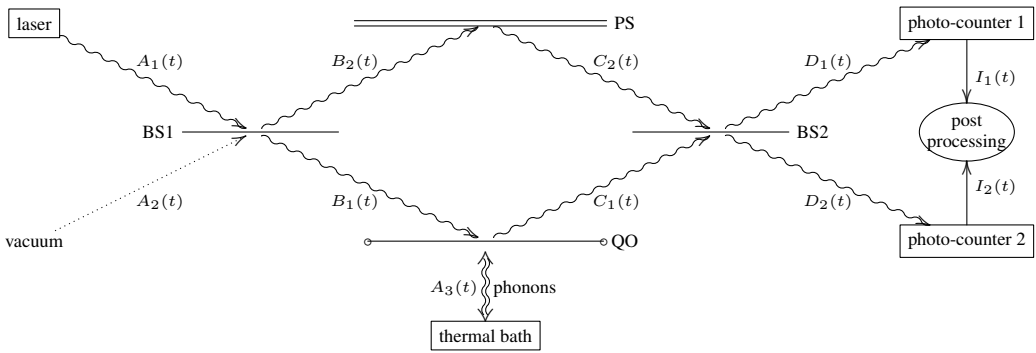

In Sect.\ \ref{Sect:oc} we introduce the Bose fields involved in QSC and the Weyl operators and we show how to describe the MZI by these objects. Let us stress that this is an important feature of the formalism based on Bose fields: very general optical circuits can be modeled in a similar way. The input light at one of the input ports of the MZI is monochromatic coherent light, while no light enter the other port. Then, we introduce the quantum observables representing  the possible detection schemes at the two output ports and we formalize the connection between the detection results and squeezing by introducing suitable mode operators. In Sect.\ \ref{Sect:HPeq} we introduce the HP-equation giving the unitary evolution of the quantum mechanical oscillator interacting with the optical field and thermal noise, with arbitrary noise spectrum. In the Heisenberg description  position and momentum of the mechanical oscillator satisfy a couple of quantum Langevin equations, which preserve  in time the canonical commutation relations, due to the unitarity of the underlying dynamics. The system operators in the HP-equation are chosen in such a way that all the forces (mechanical harmonic force, damping and radiation pressure forces) appear only in the equation for the momentum, as it must be for a mechanical oscillator. The Langevin equations can be explicitly solved and from this solution the form of the output optical field can be obtained. The explicit form of the output field for long times, together with the results about the detection schemes are used in Sect.\ \ref{Sect:det} to get the analytic expressions of the Mandel $Q$-parameters or the intensity spectra, depending on the applied measurement procedures. In the limit of weak radiation pressure interaction and of strong laser a very intense squeezing is produced and it can be detected either by the spectrum of the difference of the photo-currents, either by direct detection at the two output ports. Conclusions and possible extensions are discussed in Sect.\ \ref{sec:concl}.

\section{The optical circuit and the detection schemes}\label{Sect:oc}

Before discussing our optical circuit, the MZI in Figure \ref{MZI}, we introduce the quantum fields used to represent both light and thermal noise. A good presentation of QSC and of the related notions is in the book \cite{Parthas92};  presentations more aimed to applications in quantum optics can be found in \cite{ZolG97,Bar06}.

Let us consider $d$ Bose fields $a_j(t)$ satisfying the canonical commutations rules (CCRs)
\begin{equation}\label{CCR}
[a_i(s), a_j(t)]=0, \qquad [a_i(s), a_j^\dagger(t)]=\delta_{ij}\delta(t-s).
\end{equation}
This kind of fields are used in quantum optics also outside QSC \cite{ZGVit15}.
We work in the Fock representation, which means that these CCRs are realized in  the Hilbert space
\begin{equation}\label{Focksp}
\Gamma\equiv \Gamma\big(L^2({\mathbb R}; \Cbb^d)\big)= \Cbb\oplus \sum_{n=1}^\infty L^2({\mathbb R}; \Cbb^d)^{\otimes_s n};
\end{equation}
$\Gamma$ is the \emph{symmetric Fock space} over the one-particle space $L^2({\mathbb R}; \Cbb^d)\equiv L^2(\mathbb R)\otimes \Cbb^d $ and the direct sum on the right is its decomposition in the $n$-particle spaces.
In all developments, a key role is played by the \emph{coherent vectors}, or normalized \emph{exponential vectors}, which can be introduced by giving their components in the $n$-particle spaces:
\begin{eqnarray}\nonumber
e(f) &=&\rme^{
-\frac{1}{2}\,\norm{f}^2}\Bigl(1,f,(2!)^{-1/2}f\otimes f,\\ &{}&\ldots,(n!)^{-1/2} f^{
\otimes n},\ldots \Bigr), \qquad f\in L^2({\mathbb R}; \Cbb^d).\label{expV}
\end{eqnarray}
These vectors are completely analogous to the coherent vectors of the case of discrete modes, as one sees by comparing the representations in the spaces with fixed number of photons. Note that $e(0)$ represents the vacuum state and that we have
\begin{equation}\label{ae(f)}
a_j(t)e(f)=f_j(t)e(f);
\end{equation}
this equation  shows that $e(f)$ is indeed a coherent vector for the  introduced family of annihilation operators.

To develop the theory of quantum stochastic differential equations the integral version of the $a_j$-fields is needed, together with the integral of quadratic expressions preserving the number of quanta:
\begin{equation}\label{fdensity}
A_j(t)=\int_0^t a_j(s)\rmd s, \qquad \Lambda_{ij}^A(t)=\int_0^t a^\dagger_i(s)a_j(s)\rmd s.
\end{equation}
The operators $\Lambda_{ij}^A(t)$ were named \emph{gauge process}; note that $\Lambda_{jj}^A(t)$ is the \emph{number process} for the field $j$. The rigorous definition of field and gauge operators is through their action on the exponential vectors \cite{Parthas92}.

\subsection{The Mach-Zehnder interferometer}\label{sec:MZI}
A MZI, \cite[Sects.\ 2.4, 2.5]{WisM10}, \cite[Sect.\ 3.2.3]{HR06}, is a phase sensitive optical circuit  characterized by the presence of two beam splitters (BS) and  two mirrors.  MZI-like configurations are well suited for generation and detection of squeezed light, as in the case of \cite{HKG19}, where the squeezing is produced by non linear crystals.

Our proposal
is a Mach-Zehnder interferometer in which one of the mirrors is a reflecting and vibrating quantum oscillator (QO --- a quantum optomechanical micro-mirror), see Figure \ref{MZI}. The other mirror is a fixed one to which a tunable phase shifter (PS) has been added.
The fields $A_1$ and $A_2$ are the optical fields entering the two input ports of the interferometer, while the field $A_3$ will be used to model the thermal reservoir affecting the oscillating mirror. For the fields we are using the capital letters of the integral notation \eqref{fdensity}. In going from left to right the change of letter (from $A$ to $B$, $C$, $D$) denotes that the fields have undergone a unitary transformation. Detection applies only to the fields $D$ at the two output ports; by $I_j(t)$ we denote the two photo-currents from the detectors. The fields $D $ are detected and processed
in order to study the properties of the field $C_1$. The notion of Weyl operator, needed to describe the linear optical elements, is recalled in Sect.\ \ref{lopt+W}.
For simplicity, we assume that there are no losses and that the optical paths in the two arms of the interferometer are equal.

The idea of substituting the two mirrors in a MZI was proposed also in \cite[Sect.\ 6.5.1]{BM16}; in that reference the two optomechanical systems are coupled to cavities and the aim is to entangle two macroscopic bodies.

\subsubsection{First beam splitter}\label{Bfield}

The first beam splitter, denoted by BS1 in Figure \ref{MZI},  has transmittance $\eta \in (0,1)$ and it is represented by the Weyl operator $\Wcal_{BS1}:=\mathcal{W}(0;V_\eta)$ (see \eqref{Weyl}, \eqref{Vbs}) acting on the two optical fields with the transformation
\begin{subequations}\label{BBS1}
\begin{eqnarray}
B_j(t)&=& \Wcal_{BS1}^\dagger A_j(t)\Wcal_{BS1},
 \\
B_1(t)&=&\sqrt {\eta  }\, A_1(t)+\rmi \sqrt{1-\eta  }\, A_2(t),
\\
B_2(t)&=&\rmi \sqrt{1-\eta  }\, A_1(t)+\sqrt {\eta  }\, A_2(t).
\end{eqnarray}
\end{subequations}
Equivalently, by using the field densities, we have
\begin{eqnarray}\nonumber
b_j(t)=\sum_{i=1}^2 (V_\eta)_{ji}a_i(t), \qquad (V_\eta)_{11}&=&(V_\eta)_{22}=\sqrt {\eta  }, \\ (V_\eta)_{12}=(V_\eta)_{21}&=&\rmi \sqrt{1-\eta}.
\label{bBS1}\end{eqnarray}
The $a$-fields satisfy the canonical commutation relations (CCRs) \eqref{CCR}; being \eqref{BBS1} a unitary transformation, the same CCRs hold for the $b$-field densities.

\subsubsection{The mirrors}\label{Cfields}

\paragraph{The fixed mirror and the tunable phase shifter.} In the upper arm we have a fixed mirror, just to change the direction of propagation of the beam, and a tunable phase shifter; their effect on the beam is represented by the simple Weyl operator
\[\Wcal_{PS}=\Wcal(0;V_{PS}), \qquad
V_{PS}=\begin{pmatrix}1&0&0\\ 0&\rme^{\rmi \psi}&0 \\ 0&0&1\end{pmatrix},
\]
whose action reduces to
\begin{equation}
C_2(t)=\Wcal_{PS}^\dagger B_2(t)\Wcal_{PS}=\rme^{\rmi \psi}B_2(t), \quad \Lambda_{22}^C(t)= \Lambda_{22}^B(t),
\end{equation}
summarized in $c_2(t)=\rme^{\rmi\psi}b_2(t)$.

\begin{remark} By default we consider the phase shift to be \emph{tunable}, once for all. However, by using an electro-optical phase modulator, the phase shift $\psi$ could become time dependent and could be controlled by taking into account the detected signal (closed loop feedback) \cite[pp.\ 84, 221]{WisM10}.
\end{remark}

\paragraph{Interaction with the quantum system.} The interaction of the optical field $B_1(t)$ with the quantum system QO is represented by a Hudson-Parthasarathy equation \cite{Parthas92}. Then, the field after the interaction is given by the \emph{output field} \cite{ZolG97,Bar06,BarG13}:
\begin{subequations}\label{C=out}
\begin{eqnarray}
C_1(t) &=& B_1^{\rm out}(t):=U(t)^\dagger B_1(t)U(t),  \\ \Lambda_{11}^C(t)&=&\Lambda_{11}^{B,\,{\rm out}}(t):= U(t)^\dagger \Lambda_{11}^B(t)U(t).
\end{eqnarray}
\end{subequations}
For the moment, $U(t)$ is the unitary evolution solving a generic HP-equation. It is important to stress that HP-equation treats the quantum sub-system and the fields, representing the thermal bath and/or the light, as a closed system; then, the output fields represent the field operators in the Heisenberg picture. Note that, in such a general framework, the quantum system is not necessarily an oscillating micromirror; for example, it could be another fixed mirror, or some two-level atom absobing and emitting photons. In Section \ref{Sect:HPeq} we shall complete our model by choosing both the system interacting with the fields and their specific interactions.

We shall use also the density $c_1(t)\equiv b_1^{\rm out}(t)$. The explicit expression of this field depends on the form of the HP-equation and in our case it will be given in Sect.\ \ref{sec:IO}.

\begin{remark}\label{rem:Tt} An important point is that the fields $c_1$ and $c_2$ satisfy the CCRs as Bose free fields (cf.\ \eqref{CCR}). This is due  to the general properties of HP-equation \cite{Bar06,ZolG97,BarG13}, which imply $U(t)^\dagger B_1(t)U(t)=U(T)^\dagger B_1(t)U(T)$, for any choice of $T\geq t$. So, the output fields are obtained by a unique unitary transformation on the input fields and the commutation relations are preserved.
\end{remark}

\begin{remark} We shall consider also the case of  a fixed mirror; in this case we have
\[
C_1(t) = \rme^{\rmi\phi}B_1(t), \qquad \Lambda_{11}^C(t)=  \Lambda_{11}^B(t).
\]
\end{remark}

The two phases $\psi$ and $\phi$ take into account all the possible phase shifts induced by the MZI: reflections, beam splitters,\ldots

\subsubsection{Second beam splitter}\label{Sect:BS2}
The second beam splitter, denoted by BS2 in the figure,  has transmittance $1/2$ and it is represented by the Weyl operator $\Wcal_{BS2}=\Wcal(0;V_{1/2})$, cf.\ \eqref{BBS1}, \eqref{bBS1}, \eqref{Weyl}, \eqref{Vbs}; it generates the field transformations
\begin{subequations}\label{DC}
\begin{eqnarray}
D_j(t)&=&\Wcal_{BS2}^\dagger C_j(t) \Wcal_{BS2},\\ D_1(t)&=&\frac 1 {\sqrt 2}\left[C_1(t)+\rmi  C_2(t)\right],
\\
D_2(t)&=&\frac 1 {\sqrt 2}\left[\rmi  C_1(t)+ C_2(t)\right].
\end{eqnarray}
\end{subequations}
Equivalently, we have
\begin{equation}\label{d:cb}
d_j(t)=\frac{\rmi^{j-1}}{\sqrt 2}\left[c_1(t)-(-1)^{j}\rmi \rme^{\rmi\psi}b_2(t)\right], \quad j=1,2.
\end{equation}

\begin{remark} Again the $d$-fields are Bose free fields and satisfy the CCRs, see \eqref{d:ccr}.
\end{remark}

In the detection schemes of Sect.\ \ref{subsect:det}, a central role will be played by the number operators of the $D$-fields,
\begin{equation}\label{N:op}
\hat N_j(t)\equiv \Lambda_{jj}^D(t)=\int_0^t\rmd s\,d_j^\dagger(s)d_j(s).
\end{equation}
By \eqref{d:cb}, we have
\begin{eqnarray}\nonumber
\hat N_j(t)&=&\frac 12 \int_0^t\rmd s\Bigl[c_1^\dagger(s)c_1(s)
-(-1)^j \rmi \rme^{\rmi\psi} c_1^\dagger(s)b_2(s)\\ &&{}+(-1)^j \rmi \rme^{-\rmi\psi}b_2^\dagger(s) c_1(s)+b_2^\dagger(s)b_2(s)\Bigr].
\label{DCjj}\end{eqnarray}

\subsection{The total system state}\label{Sect:fs}
Our total system is composed by the quantum oscillator, the two optical fields, $A_1,\; A_2$, and a thermal field $A_3$ interacting directly with the oscillator. As initial state we take a factorized state composed by a generic statistical operator $ \rho_\rmm^0$ for the oscillator, a coherent state for the field  $A_1$, the vacuum for the field $A_2$, and a thermal state for $A_3$.

The general form of the coherent states for the fields is given in \eqref{expV}; they are characterized by  $L^2(\Rbb)$-functions. To have the vacuum for the field $A_2$ it is enough to take the null function, but to  introduce a monochromatic laser for the field $A_1$ we need to have a starting and a ending time for the function. So, we take
\begin{subequations}\label{fT}
\begin{eqnarray}
\rho_T&=&\rho_\rmm^0 \otimes \rho_{\rm em}^{T} \otimes \rho_{\rm th}^T, \qquad \rho_{\rm em}^{T}= \rho_{{\rm em}1}^{T} \otimes \rho_{{\rm em}2},
\\ \rho_{{\rm em}1}^{T}&=& |e_1(f_T)\rangle \langle e_1(f_T)|,\quad f_T(t)= f(t)\ind_{(0,T)}(t), \nonumber
\\ f(t)&=& \lambda\rme^{-\rmi \omega_0 t}, \quad \lambda \in \Cbb, \quad \omega_0>0,
\\  \rho_{{\rm em}2}&=&|e_2(0)\rangle \langle e_2(0)|.
\end{eqnarray}
\end{subequations}
In all the following developments, the current time $t$ will be always smaller than $T$, but in the final physical formulae we shall take $T\to +\infty$.

Also mixture of coherent states could be used, such as
the phase-diffusion model of a laser \cite{Bar16}; however, due to the fact that we consider only the case of equal optical paths in the two arms of the MZI the only property of the field state we shall use is $\abs{f(t)}^2=\abs\lambda^2$; so, the pure monochromatic case is sufficient.

The state $\rho_{\rm th}^T$ will be discussed in Sect.\ \ref{sec:tfst}.

We shall denote the quantum expectation of any operator $X$ by the notation
\begin{equation}
\langle X\rangle_T\equiv \Tr\{X\rho_T\}.
\end{equation}
By using \eqref{d:cb}, we get the following useful formula, which enables to compute field moments,
\begin{equation}\label{d:Psi}
d_j(t)\, \rho_{\rm em}^{T}=\frac{\rmi^{j-1}}{\sqrt 2}\left[c_1(t)+(-1)^{j} \rme^{\rmi\psi}\sqrt{1-\eta  }\,f_T(t)\right] \rho_{\rm em}^{T}.
\end{equation}

\subsection{The detection schemes}\label{subsect:det}

We describe now some possible detection schemes. The first one is to consider the simple counting processes of the photons from the two output ports. Then, we shall consider possible post-processing of the output currents from the detectors and study their spectra.

The key point in the mathematical formulation of detection schemes we shall consider is that the number operators \eqref{N:op}, $\{\hat N_j(t),\ j=1,2, \ 0\leq t\leq T\}$, are a family of commuting selfadjoint operators. Indeed, from the CCRs \eqref{d:ccr} we obtain
\begin{equation}\label{Nop+[]}
\left[\hat N_i(t),\, \hat N_j(s)\right]=0, \quad \forall i,j =1,2, \quad \forall t,s\in[0,T].
\end{equation}
Because they commute, these operators represent \emph{compatible observables}, the number of photons up to time $t$. We denote by $N_j(t)$ the observed counts,  whose joint probability distribution $P$ can be in principle obtained by the ``usual'' rules of quantum mechanics (from the joint projection valued measure and the system state); they form a two-dimensional stochastic process, whose components are two counting processes. Obviously, also functions of these commuting operators represent compatible observables.

The physical quantities introduced in the following sections, like variances and spectra, can be expressed firstly by means of the fields $D_j$ and their densities $d_j$ (see the detected fields in Section \ref{Sect:BS2}). The expressions we get in this way are independent from the explicit structure of the MZI, so they are very general. Then, they are particularized to our optical circuit, but the interaction fields/optomechanical component is left completely general (see the output field in Section \ref{Cfields}). The mathematical computations are given in Appendix \ref{app:Var+sp}; these computations are essentially the same either for the case of direct detection, either in the case of the intensity spectra. After the introduction of the interaction fields/optomechanical component, given in Section \ref{Sect:motion}, we shall obtain the final explicit results in Section \ref{Sect:det}.

\subsubsection{The counting processes}\label{sec:counts}

Here we consider the case of \emph{direct detection} at the two output ports of the MZI, which means to count the photons leaving the two ports. So, the observed quantities are the processes $N_j(t)$, just introduced above, and the associated self-adjoint operators are the number operators $\hat N_j(t)$ \eqref{N:op}. In particular, the means at time $T$ are
\begin{equation}\label{meanN}
\Ebb_P[N_j(T)]=\langle \hat N_j(T)\rangle_T =\int_0^T \rmd t\, \langle d_j^\dagger(t)d_j(t)\rangle_T;
\end{equation}
in the second equality we have expressed the number operator in terms of the $d$-densities as in \eqref{DCjj}.
Due to the stationarity of the stimulating laser (see the state \eqref{fT}), there exists an asymptotic regime for the fields, and  the following limits exist
\begin{eqnarray}\nonumber
n_j&:=&\lim_{T\to +\infty}\,\frac{\Ebb_P[N_j(T)]}T\\ {}&=&\lim_{t\to+\infty}\lim_{T \to +\infty}\langle d_j^\dagger(t)d_j(t)\rangle_T, \qquad j=1,2,
\label{def:nj}\end{eqnarray}
which represent the mean flux of photons at large times. The proof of the existence of the limit is by explicit computations, see Remark \ref{rem:nj} and Appendix \ref{Dmom_comp}.

As we have observables represented by commuting self-adjoint operators, variances and covariances are given by the quantum expectations
\begin{eqnarray}\nonumber
\Cov_P&&[N_i(T),N_j(T)]\\ {}&&= \langle \hat N_i(T)\hat N_j(T)\rangle_T - \langle \hat N_i(T)\rangle_T\langle \hat N_j(T)\rangle_T.
\label{CovNN}\end{eqnarray}
Then, by inserting the integral representation of the number operators and by using \eqref{jjii:jiij},  we obtain the expression
\begin{eqnarray}\nonumber
\Cov_P&&[N_i(T),N_j(T)]=\delta_{ij}\langle \hat N_j(t)\rangle_T
\\ \nonumber &&{}+\int_0^T \rmd t \int_0^T\rmd s
\Bigl(\langle d_j^\dagger(t)d_i^\dagger(s)d_i(s)d_j(t)\rangle_T
\\ &&{}- \langle d_i^\dagger(s)d_i(s)\rangle_T\langle d_j^\dagger(t)d_j(t)\rangle_T \Bigr).
\label{CovNN1}\end{eqnarray}

In \eqref{meanN}--\eqref{CovNN1}, the mean rates of counts, their variances and covariance are given  in terms of the detected field densities $d_j$. The compatibility of the considered observables and the structure of all these formulae  are due only to the CCRs \eqref{d:ccr} obeyed by the $d$-fields; all these results are independent from the specific structure of the optical circuit.

In the case of Poisson distribution the variance of the observable $N_j$ would be equal to its mean. In quantum optics it is usual to measure the difference from this distribution by introducing the \emph{Mandel $Q$-parameter} \cite{M79}:
\begin{equation}\label{MQp}
Q_j(T):=\frac{\Var_P[N_j(T)]- \Ebb_P[N_j(T)]}{\Ebb_P[N_j(T)]};
\end{equation}
by definition the $Q$-parameter is greater than $-1$ and vanishes for the Poisson distribution.
As for \eqref{def:nj}, also for this quantity it will exist the limit for large times in our case; by \eqref{CovNN1} we can write
\begin{eqnarray}\nonumber
Q_j&&:=\lim_{T\to +\infty} Q_j(T)
\\ \nonumber {}= &&\lim_{T\to +\infty}\frac 1{n_jT}\int_0^T \rmd t \int_0^T\rmd s
\Bigl(\langle d_j^\dagger(t)d_i^\dagger(s)d_i(s)d_j(t)\rangle_T
\\{}&&- \langle d_i^\dagger(s)d_i(s)\rangle_T\langle d_j^\dagger(t)d_j(t)\rangle_T \Bigr)\geq-1.
\label{Qinfty}\end{eqnarray}
When the Mandel parameter is negative, one speaks of sub-Poissonian statistics and this fact has connections with the presence of squeezed light
\cite[Sect.\ 2.3.6]{Car99}; we shall discuss this point in Sect.\ \ref{sect:squeeze}.

\subsubsection{Post-processing and the spectra of the output light}\label{sec:pp&det}
We consider now the case of photo-counters designed to measure the flux of arrivals of photons. The photo-detector does not distinguish the single detected photons, but  it produces an output photo-current proportional to a time mean of the electrical pulses generated by the incoming photons. We consider a simple explicit representation of the \emph{response function} of the detector given by $c\varkappa\rme^{-\varkappa \left(t-r\right)}$,  so that the output currents are represented by
\begin{eqnarray}\label{Ij(t)}
&&I_j(t) = c\varkappa \int_0^t \rme^{-\varkappa \left(t-r\right)}\,\rmd N_j(r), \\ &&c>0, \quad \varkappa>0, \quad j=1,2, \quad 0<t\leq T;
\nonumber\end{eqnarray}
we are considering two identical, ideal detectors.  Being $N_j$ a counting process, its infinitesimal increments have to be interpreted as 0 or 1: they are 1 at the times of a count \cite[p.\ 22]{ZolG97}; then, the stochastic integral in \eqref{Ij(t)} is a sum on the random times of the arrivals of the photons in the time interval $(0,t)$. The associated operators are given by
\[
\hat I_j(t) = c\varkappa \int_0^t \rme^{-\varkappa \left(t-r\right)}\,\rmd \Lambda_{jj}^D(r).
\]
As we already remarked, the number operators  form a family of commuting self-adjoint operators; so, the same holds for the family of current operators $\left\{\hat I_1(t),\, \hat I_2(s)\right\}_{t,s\in[0,T]}$.
We can also process the output currents and measure the ``sum'' current or the ``difference'' current:
\begin{equation}\label{Ipm}
I_+(t)=I_1(t)+ I_2(t), \qquad I_-(t)=I_1(t)- I_2(t);
\end{equation}
the corresponding commuting self-adjoint operators will be denoted by  $\hat I_\pm(t)=\hat I_1(t)\pm \hat I_2(t)$. A special role is played by $I_-(t)$, which is the quantity measured in the balanced homodyne detection scheme \cite{GarZ00,WisM10,Leo10,Yur85}. Indeed, it is the MZI itself which realizes the circuit needed for homodyning. From  Figure \ref{MZI} one sees that the field $C_2(t)$ plays the role of local oscillator, while $C_1(t)$ is the signal to be analyzed. They interfere at the balanced beam splitter BS2; then, the fields $D_j(t)$ are detected, the photocurrents are subtracted and the resulting signal is processed.

Anyone of the currents \eqref{Ij(t)} or \eqref{Ipm} can be sent to a spectrum analyzer and its intensity spectrum observed. Let us denote by
$I(t)$ one of these currents or a linear combination of them, $I(t) = \sum_{j=1}^2 k_jI_j(t)$, $k_j\in \Rbb$,
and let $\hat I(t)$ be the corresponding family of compatible quantum observables; its intensity spectrum is given by
\begin{subequations}\label{SI*}
\begin{eqnarray}
S_{I}^T(\mu)&=&\frac 1T\,\Ebb_P\left[\abs{\int_0^T \rme^{\rmi \mu t}I(t)\rmd t}^2\right], \\ S_{I}(\mu)&=&\lim_{T\to +\infty}S_{I}^T(\mu).
\end{eqnarray}
\end{subequations}
This is the usual definition of  spectrum of an asymptotically stationary stochastic process, as it is in our case, see Sect.\ \ref{Sect:det}  and the computations in Appendix \ref{Dmom_comp}. The limit is in the sense of distributions, as Dirac-deltas can appear in the limit.

\begin{remark}\label{SimmS} The intensity spectra are  symmetric. Indeed, as $I$ is a real stochastic process, from the definition we have
\begin{eqnarray*}
S_{I}^T(-\mu)&=&\frac 1T\,\Ebb_P\left[\int_0^T \rme^{-\rmi \mu t}I(t)\rmd t \int_0^T \rme^{\rmi \mu s} I(s)\rmd s\right] \\ {}&=&S_{I}^T(\mu).
\end{eqnarray*}
\end{remark}

By expressing the probabilistic means as quantum expectations, firstly
we get from \eqref{def:nj} the mean of the current at large times:
\begin{eqnarray}\nonumber
\lim_{t\to+\infty}\Ebb_P[I(t)]&=&\lim_{t\to+\infty}\lim_{T\to+\infty}\langle \hat I(t)\rangle_T \\ {}&=&c\left(k_1n_1+k_2n_2\right).
\end{eqnarray}
Then, by subtracting and adding the contribution of the means in \eqref{SI*}, we obtain
\begin{eqnarray}\nonumber
S_{I}(\mu)&=&2\pi c^2 \left(k_1n_1+k_2n_2\right)^2 \delta (\mu)\\ {}&+& \frac{c^2\varkappa^2}{\mu^2 +\varkappa^2}\sum_{ij}k_i\left[ n_i\delta_{ij}+\abs\lambda^2\Sigma_{ij}(\mu)\right]k_j,
\label{SI*2}\end{eqnarray}
\begin{eqnarray}\nonumber
\Sigma_{ji}(\mu)&:=&\lim_{T\to +\infty}\frac 1{\abs\lambda^2T}\,\RE\int_0^T \rmd t \int_0^T \rmd s \,\rme^{\rmi\mu \left(t-s\right)}
\\ \nonumber {}&\times& \Bigl(\langle d_j^\dagger(t)d_i^\dagger(s) d_i(s)d_j(t)\rangle_T
\\ {}&-&\langle d^\dagger_j(t)d_j(t)\rangle_T\, \langle d^\dagger_i(s)d_i(s)\rangle_T\Bigr);
\label{Sigmaji}\end{eqnarray}
the proof is given in Appendix \ref{dj:S}.  These expressions of the intensity spectra are independent from the specific structure of the optical circuit, as it was the case of the covariance \eqref{CovNN1}.

\begin{remark}\label{rem:spdec} In the expression \eqref{SI*2}, the term with the Dirac delta is the contribution of the constant component of the currents. The factor $\frac{c^2\varkappa^2}{\mu^2 +\varkappa^2}$ is the square modulus of the Fourier transform of the response function of the detector. The term $\delta_{ij}n_j$ is known as \emph{shot noise} \cite[p.\ 362]{GarZ00}; it derives from having written the expression inside \eqref{Sigmaji}  in normal order by using \eqref{jjii:jiij}.
Finally, the matrix $\left[ n_i\delta_{ij}+\abs\lambda^2\Sigma_{ij}(\mu)\right]$ is positive semi-definite by construction.
\end{remark}

By comparing \eqref{Sigmaji} with \eqref{CovNN1}, we see that we have the connection
\begin{equation}\label{CovNN2}
\lim_{T\to+\infty}\frac 1T\,\Cov_P[N_i(T),N_j(T)]=\delta_{ij}n_j+\abs\lambda^2\Sigma_{ji}(0).
\end{equation}

\subsubsection{The dependence on the output field}\label{sec:c1out}
Our interest is on the field $C_1(t)$, the optical field after the interaction with the opto-mechanical device. By comparing our optical circuit with a \emph{balanced homodyne detection} scheme, see for instance \cite[Sect.\ 8.4.4 and Fig.\ 8.7]{GarZ00}, we note that there is a strict similarity and that the quantum field $C_2$ has the role of the \emph{local oscillator} in the homodyne scheme. So, we put in evidence here the dependence of the observed quantities on the field density $c_1(t)$.

By using \eqref{fT} and \eqref{d:Psi},  we get easily from \eqref{def:nj} the following expression for the mean rate at port $j$: \footnote{``c.c.'' in an equation means ``complex conjugated terms''.}
\begin{eqnarray}\nonumber
n_j&=&\lim_{t\to+\infty} \lim_{T\to+\infty} \frac12 \Bigl\{\abs\lambda^2\left(1-\eta\right)+\langle c_1^\dagger(t) c_1(t)\rangle_T
\\ {}&+& (-1)^j\sqrt{1-\eta} \left(\rme^{\rmi \psi} f(t)\langle  c_1^\dagger(t)\rangle_T
 +\text{c.c.}\right)\Bigr\}.
\label{meanN3}\end{eqnarray}

We introduce now the \emph{reduced spectra}, from which all our physical quantities can be expressed: \footnote{The overline denotes the complex conjugation.}
\begin{subequations}\label{Sigmas}
\begin{equation}\label{Sigma-}
\Sigma_-(\mu) =\Sigma_-^0(\mu)+\Sigma_-^\psi(\mu),
\end{equation}
\begin{eqnarray}\nonumber
\Sigma_-^0(\mu) &=& \lim_{T\to+\infty}\frac {1} {\abs\lambda^2 T}\int_0^T \rmd t \int_0^T \rmd s \, \rme^{\rmi\mu \left(t-s\right)}
\\ \nonumber{}&\times& \Bigl\{\overline{f(t)}\,f(s) \langle c_1^\dagger(s)c_1(t)  \rangle_T
\\ {}&-&\overline{f(t)}\,f(s)\langle c_1^\dagger(s)\rangle_T  \langle c_1(t)  \rangle_T+ \text{c.c.}\Bigr\},
\label{Sigma-0}\end{eqnarray}
\begin{eqnarray}\nonumber
\Sigma_-^\psi&&(\mu) = \lim_{T\to+\infty}\frac {1} {\abs\lambda^2 T}\int_0^T \rmd t \int_0^T \rmd s \, \rme^{\rmi\mu \left(t-s\right)}
\Bigl\{ \rme^{2\rmi \psi}f(s)
\\ {}\times&& f(t)
\Bigl[\langle c_1^\dagger(t)  c_1^\dagger(s)\rangle_T  -\langle c_1^\dagger(s)\rangle_T \langle c_1^\dagger(t)  \rangle_T \Bigr]
+ \text{c.c.}\Bigr\},
\label{Sigma-psi}\end{eqnarray}
\begin{eqnarray}\nonumber
\Sigma_0&& (\mu) = \lim_{T\to+\infty}\frac 1 {\abs\lambda^2 T}\int_0^T \rmd t \int_0^T \rmd s \, \rme^{\rmi\mu \left(t-s\right)}
\\ \nonumber &&{}\times
\Bigl[ \rme^{\rmi \psi}
\Bigl( f(t)\langle c_1^\dagger(t)  c_1^\dagger(s)c_1(s) \rangle_T
-f(t)\langle c_1^\dagger(t)\rangle_T
\\ \nonumber &&{} \times  \langle c_1^\dagger(s)c_1(s) \rangle_T + f(s)\langle  c_1^\dagger(s)c_1^\dagger(t) c_1(t)\rangle_T
\\ && {}-f(s)\langle  c_1^\dagger(s)\rangle_T \langle c_1^\dagger(t) c_1(t)\rangle_T \Bigr)
+ \text{c.c.}\Bigr],
\label{Sigma0}\end{eqnarray}
\begin{eqnarray}\nonumber
\Sigma_+(\mu) &=& \lim_{T\to+\infty}\frac 1 {\abs\lambda^2 T}\int_0^T \rmd t \int_0^T \rmd s \, \rme^{\rmi\mu \left(t-s\right)}
\\ \nonumber {}&\times& \Bigl( \langle c_1^\dagger(t)  c_1^\dagger(s)c_1(s)c_1(t)\rangle_T
\\ {}&-&\langle c_1^\dagger(t)c_1(t)\rangle_T\, \langle  c_1^\dagger(s)c_1(s)\rangle_T\Bigr).
\label{Sigma+}\end{eqnarray}
\end{subequations}
As discussed in Appendix \ref{sec:c1} they are real and symmetric: $\Sigma_{\pm}(-\mu)=\Sigma_\pm(\mu)$, \quad $\Sigma_{0}(-\mu)=\Sigma_0(\mu)$. Moreover, from the explicit structures \eqref{Sigma-0} and \eqref{Sigma-psi}, we see that
\begin{equation}\label{90b}
\Sigma_-^0(\mu)\geq 0, \qquad \Sigma_-^\psi(\mu)+\Sigma_-^{\psi+\pi/2}(\mu)=0.
\end{equation}

In Appendix \ref{sec:c1} it is shown that the components \eqref{Sigmaji} of the spectra can be expressed by means of \eqref{Sigmas} in the following way:
\begin{subequations}\label{Sigma:c1}
\begin{eqnarray}\nonumber
\Sigma_{jj}(\mu) = \frac 14 &\Bigl[&\Sigma_+(\mu) + (-1)^j \sqrt{1-\eta  }\,\Sigma_0(\mu)\\ {} &+&\left(1-\eta\right)\Sigma_-(\mu)\Bigr],
\label{Sjj3}\end{eqnarray}
\begin{equation}
\label{S12:cov3}
\Sigma_{12}(\mu) =\frac 14\left[\Sigma_+(\mu) -\left(1-\eta\right)\Sigma_-(\mu)\right].
\end{equation}
\end{subequations}
Then, by inserting these expressions into \eqref{SI*2} we can derive the four spectra which we are interested in:
\begin{eqnarray}\nonumber
S_{I_+}(\mu) &=&2\pi c^2\left(n_1+n_2\right)^2\delta(\mu)\\ {}&+&\frac{c^2\varkappa^2}{\mu^2 +\varkappa^2}\left[n_1+n_2+\abs\lambda^2\Sigma_+(\mu)\right],
\label{S+}\end{eqnarray}
\begin{eqnarray}\nonumber
&&S_{I_-}(\mu) =2\pi c^2\left(n_1-n_2\right)^2\delta(\mu)\\ {}&&\qquad{}+\frac{c^2\varkappa^2}{\mu^2 +\varkappa^2}\left[n_1+n_2+\abs\lambda^2\left(1-\eta\right)\Sigma_-(\mu)\right],
\label{S-}\end{eqnarray}
\begin{eqnarray}\nonumber
S_{I_j}(\mu) &=&2\pi c^2n_j^2\delta(\mu)+\frac{c^2\varkappa^2}{\mu^2 +\varkappa^2}\biggl\{n_j +\frac {\abs\lambda^2}4 \Bigl[\Sigma_+(\mu)
\\ {}&+& (-1)^j \sqrt{1-\eta  }\,\Sigma_0(\mu) +\left(1-\eta\right)\Sigma_-(\mu)\Bigr]\biggr\}.
\label{Sj}\end{eqnarray}
The meaning of the various terms is just as in Remark \ref{rem:spdec}; note that the shot noise is the same for the spectra of the two currents $I_\pm$.

Now, let us go back to the direct detection and the counting processes introduced in Section \ref{sec:counts}. We have already obtained the mean rates in terms of the $c_1$-field, see Eq.\ \eqref{meanN3}. From \eqref{MQp}, \eqref{Qinfty}, \eqref{CovNN2} we obtain the  asymptotic $Q$-parameters for the counting processes at the two output ports:
\begin{equation}
Q_j= \frac {\abs\lambda^2} { 4n_j}  \Bigl[ \Sigma_+(0)+(-1)^j\sqrt{1-\eta}\,\Sigma_0(0)
+\left(1-\eta\right)\Sigma_-(0)
\Bigr].
\label{MQpj}\end{equation}
We can consider also the counting process ``sum'' and the point process ``difference''; their variances can be computed from the usual probabilistic identity
\begin{eqnarray*}
\Var_P[N_1(T)\pm N_2(T)]&=& \Var_P[N_1(T)]+ \Var_P[N_2(T)]
\\ {}&\pm& 2 \Cov_P[N_1(T),N_2(T)];
\end{eqnarray*}
their long time behaviour  turns out to be
\begin{equation}
\lim_{T\to+\infty}\frac{\Var_P[N_1(T)+ N_2(T)]}T
=n_1+n_2+\abs\lambda^2 \Sigma_+(0),
\label{Var+/T}\end{equation}
\begin{eqnarray}\nonumber
\lim_{T\to+\infty}&&\frac{\Var_P[N_1(T)- N_2(T)]}T
\\ &&{}=n_1+n_2
+\abs\lambda^2 \left(1-\eta\right)\Sigma_-(0).
\label{Var-/T}\end{eqnarray}
Then, the Mandel $Q$-parameter of the counting process ``sum'' is
\begin{eqnarray}\nonumber
Q_+&:=&\lim_{T\to +\infty} \frac{\Var_P[N_1(T)+N_2(T)]- \sum_{j=1}^2\Ebb_P[N_j(T)]} {\sum_{i=1}^2\Ebb_P[N_i(T)]}
\\ {}&=&\frac{\abs\lambda^2}{n_1+n_2}\, \Sigma_+(0).
\label{MQp+}\end{eqnarray}
The difference of counts is not a counting process; however, by analogy, we can introduce a $Q$-parameter also for this case:
\begin{eqnarray}\nonumber
Q_-&:=&\lim_{T\to +\infty} \frac{\Var_P[N_1(T)-N_2(T)]- \sum_{j=1}^2\Ebb_P[N_j(T)]}{\sum_{i=1}^2\Ebb_P[N_i(T)]}
\\ {}&=&\frac{\abs\lambda^2}{n_1+n_2} \left(1-\eta\right)\Sigma_-(0).
\label{MQp-}\end{eqnarray}

\subsubsection{Shot noise reduction and squeezed light}\label{sect:squeeze}
The presence of shot noise reduction is usually attributed to squeezed light \cite{Car08,WalM94}. Here we formalize this connection following
\cite{BarG13}, where a general balanced homodyne detection is analyzed inside the theory of measurements in continuous time.

By the gated Fourier transfom of the field $c_1(t)$, introduced in Sect.\ \ref{Cfields}, we define its frequency components
\begin{eqnarray}\nonumber
c_T(\mu)&:=& \frac 1{\abs\lambda\sqrt T}\int_0^T\rmd t\,\rme^{\rmi\mu t}\,\overline{f(t)}\, c_1(t)
\\{}&=& \frac \lambda{\abs\lambda\sqrt T}\int_0^T\rmd t\,\rme^{\rmi\left(\mu +\omega_0\right) t}\, c_1(t).
\label{cmu}\end{eqnarray}
It is easy to check that the following commutations rules hold:
\begin{equation}
[c_T(\mu), c_T^\dagger(\mu')]=\begin{cases} 1 & \text{for} \ \mu=\mu' ,
\\ \frac{\rme^{\rmi\left(\mu-\mu'\right)T}-1}{\rmi\left(\mu-\mu'\right)T}& \text{for} \ \mu\neq \mu'.
\end{cases}
\end{equation}
By this, $c_T(\mu)$ is a ``mode operator''; however, for $\mu\neq \mu'$, $c_T(\mu)$ and $c_T(\mu')$ do not commute and, so, they do not act on orthogonal Hilbert subspaces. We can say that they become approximately orthogonal  for $\abs{\mu-\mu'}T\to +\infty$. The operators \eqref{cmu} are examples of the \emph{temporal filtered modes} discussed in \cite[Sect.\ 2.1]{ZGVit15}.

Then, we define the ``two-modes quadrature'' operators $Q_T(\mu;\psi)$ and their fluctuation parts $\Delta Q_T(\mu;\psi)$ by
\begin{eqnarray}\label{Qmu}
Q_T(\mu;\psi)&:=&\rme^{\rmi\psi}c_T(\mu) + \rme^{-\rmi\psi}c_T^\dagger(-\mu),
\\ \label{DeltaQmu}
\Delta Q_T(\mu;\psi)&:=&Q_T(\mu;\psi)- \langle Q_T(\mu;\psi)\rangle_T ;
\end{eqnarray}

By using \eqref{cmu} into \eqref{Qmu} we obtain \footnote{``h.c.'' in an equation means ``hermitian conjugated terms''.}
\begin{equation*}
Q_T(\mu;\psi)=\frac 1 T\int_0^T \rmd t \,\rme^{\rmi \mu t} \left(\frac \lambda {\abs{\lambda}}\, \rme^{\rmi \left(\psi+\omega_0t\right)}c_1(t)+\text{h.c.}\right),
\end{equation*}
from which we see that $Q_T(\mu;\psi)$ is a finite-time Fourier component of the quadrature $\frac \lambda {\abs{\lambda}}\, \rme^{\rmi \left(\psi+\omega_0t\right)}c_1(t)+\text{h.c.}$ of the field density $c_1(t)$. Similar Fourier components have been introduced in \cite[Sect.\ 9.3]{Car08}, inside the treatment of the \emph{spectrum of squeezing}. For simplicity, we have extended the term ``quadrature'' also to $Q_T(\mu;\psi)$.

The operators \eqref{Qmu} enjoy the properties:
\begin{eqnarray*}
Q_T(\mu;\psi)^\dagger &&=Q_T(-\mu;\psi),\qquad [Q_T(\mu;\psi),Q_T(\mu';\psi)]=0, \\ &&[Q_T(\mu;\psi),Q_T(-\mu;\psi\pm\pi/2)]=\mp2\rmi.
\end{eqnarray*}
Then, a Heisenberg-like relation holds for the ``fluctuation operators'' \eqref{DeltaQmu}, \cite[Theorem 7]{BarG13}:
\begin{eqnarray}\nonumber
\langle&&\Delta Q_T(\mu;\psi)^\dagger \Delta Q_T(\mu;\psi)\rangle_T\\\ &&{}\times\langle\Delta Q_T(\mu;\psi\pm\pi/2)^\dagger \Delta Q_T(\mu;\psi\pm\pi/2)\rangle_T\geq 1.\label{Heis}
\end{eqnarray}

It will be useful to have a notation for the quadrature fluctuations for large $T$:
\begin{equation}\label{Delta()}
\Delta^2(\mu,\psi):=\lim_{T\to +\infty}\langle \Delta Q_T(\mu;\psi)^\dagger \Delta Q_T(\mu;\psi)\rangle_T\geq 0.
\end{equation}
Then,
by comparing \eqref{Sigma-}--\eqref{Sigma-psi} with \eqref{cmu}, \eqref{Qmu}, \eqref{DeltaQmu}, we obtain the following representation of the reduced spectrum  in terms of the quadrature fluctuations:
\begin{equation}\label{cmurepr}
1+\Sigma_-(\mu)=\Delta^2(\mu,\psi).
\end{equation}
From \eqref{Heis} and \eqref{cmurepr}, we get the two bounds
\begin{subequations}\label{2bounds}
\begin{eqnarray}\label{UR}
\left(1+\Sigma_-^0(\mu)+\Sigma_-^\psi\right)&&\left(1+\Sigma_-^0(\mu)+\Sigma_-^{\psi+\pi/2}\right)\geq 1,
\\
\Sigma_-(\mu)&&\equiv\Sigma_-^0(\mu)+\Sigma_-^\psi \geq -1;
\end{eqnarray}
\end{subequations}
recall that $\Sigma_-^0(\mu)\geq 0$, $\forall\mu$.

By this construction we see that on a coherent vector for the modes $c_T(\mu)$ and $c_T(-\mu)$  we have $\Delta^2(\mu,\psi)=1$; in particular this is true for the vacuum state.
Following \cite{Car08} we speak of \emph{squeezing} when the fluctuations of a certain quadrature are reduced under their value in the vacuum state; by \eqref{cmurepr}, this means when $\Sigma_-(\mu)<0$ for some $\mu$ and $\psi$. Then, the Heisenberg-like relation \eqref{2bounds} says that, when  the quadrature \eqref{Qmu}
is squeezed, the complementary quadrature is anti-squeezed.

A particular role is played by the mode operator for $\mu=0$; in this case $ Q_T(0;\psi)$ involves the single mode  $c_T(0)$, $c_T^\dagger(0) $ and it is self-adjoint. To have $\Sigma_-(0)$ near the lower bound $-1$ means that the field state is near an eigenstate of $ Q_T(0;\psi)$, i.e.\ a very squeezed state (but recall that the bound cannot be reached, because $ Q_T(0;\psi)$ has a continuous spectrum). Indeed, in Sect.\ \ref{sqdetect} we shall detect squeezing by finding conditions under which $\Sigma_-(0)$ is near its lower bound. Note that squeezing can be detected also by the counting of photons, because of  the proportionality of the $Q$-parameter $Q_-$ \eqref{MQp-} to $\Sigma_-(0)$.

The monitoring schemes and the whole optical circuit discussed above are aimed  to produce and to detect squeezing in the light after the scattering on the movable mirror. In \cite{Bar16} the same interaction mirror/light was considered, but with heterodyne detection. That set up was not suitable to detect squeezing, and indeed different effects were  highlighted.

\begin{remark}[Random phases] Equations \eqref{90b}, \eqref{2bounds} recall us that squeezing depends strongly on the phase. If some phase is not stable, say $\psi$ is highly random in time, from \eqref{Sigma-}--\eqref{Sigma-psi}  we get that
$\Sigma_-(\mu)$ reduces to $\Sigma_-^0(\mu)$, which is always positive.
\end{remark}

\subsection{The MZI without the quantum oscillating mirror}\label{noQ}
To better understand the behaviour of a MZI, let us consider the case in which our oscillating quantum system is replaced by a fixed mirror introducing a phase $\phi$:\  $c_1(t)= \rme^{\rmi \phi}b_1(t)$. By using this expression in \eqref{meanN3}, \eqref{Sigmas}
we get
\begin{eqnarray}\label{Sigma=0}
n_j&=&\frac{\abs\lambda^2}2 \left[1+(-1)^j \chi\cos \left(\phi-\psi\right)\right], \\ &&\Sigma_\pm(\mu)=\Sigma_0(\mu)=0.
\end{eqnarray}
Recall that $\chi=2\sqrt{\eta \left(1-\eta  \right)}\in[0,1]$, and $\chi=1$ for $\eta =1/2$.

\paragraph{The counting processes.}
From \eqref{meanN}, \eqref{def:nj}, \eqref{Sigma=0} we get
\begin{eqnarray*}
&&\Ebb_P[N_j(T)]=\Ebb_P\langle \Lambda^D_{jj}(T)\rangle =Tn_j, \\ &&\Ebb_P[N_1(T)]+\Ebb_P[N_2(T)]=\abs\lambda^2 T.
\end{eqnarray*}

\begin{remark} For $\eta  =1/2$, both beam splitters are perfectly balanced and we get
\[
\Ebb_P[N_1(T)]=\frac 12\abs\lambda^2 T \left[1-\cos\left(\psi-\phi\right)\right].
\]
In the case $\cos\left(\psi-\phi\right)=1$, there is no light coming out from port 1; for $\cos\left(\psi-\phi\right)=-1$ there is no light from port 2. This property of a balanced MZI is well known in classical optics, see \cite[Sect.\ 3.2.3]{HR06}.
\end{remark}

From \eqref{CovNN2},  \eqref{Sigmas}, \eqref{Sigma:c1}, \eqref{MQpj}, \eqref{MQp+}, \eqref{MQp-}, we get
\begin{eqnarray}\nonumber
\Cov_P[N_1(T),N_2(T)]&=&0,
\\
\lim_{T\to +\infty}\frac 1T\,\Var_P[N_j(T)]&=&n_j,
\\ \nonumber
Q_j=0 , \qquad Q_\pm&=&0.
\end{eqnarray}

\paragraph{The spectrum of the ouput currents.}
From \eqref{Sigma:c1}--\eqref{Sj},  we get
\begin{eqnarray*}
S_{I_j}(\mu) &=& \frac{c^2\varkappa^2}{\mu^2 +\varkappa^2}\,n_j+2\pi c^2 n_j^2\delta(\mu),
\\ S_{I_+}(\mu)&=&\frac{c^2\abs\lambda^2\varkappa^2}{\mu^2 +\varkappa^2 }
+2\pi c^2 \abs{\lambda}^4\delta(\mu),
\\
S_{I_-}(\mu)
&=&\frac{c^2\abs\lambda^2\varkappa^2}{\mu^2 +\varkappa^2 }
+2\pi c^2\chi^2 \abs{\lambda}^4\delta(\mu)\,\cos^2\left(\psi-\phi\right).
\end{eqnarray*}
Only the shot-noise component and the $\delta$-component due to the constant mean contribute to the spectra. The $\delta$-peak in $S_{I_-}(\mu)$ can be made vanishing by tuning the phase, while $S_{I_+}(\mu)$ is independent from the phases and from $\eta $.

In the case $\eta =1/2$ we have $\chi=1$ and
\begin{eqnarray}\nonumber
S_{I_1}(\mu)&=&\frac{\pi c^2\abs\lambda^4}2\left[1- \cos\left(\phi-\psi\right)\right]^2\delta(\mu) \\ {}&+& \frac{c^2\abs\lambda^2\varkappa^2} {2(\mu^2+\varkappa^2)}\left[1- \cos\left(\phi-\psi\right)\right],
\\ \nonumber
S_{I_2}(\mu)&=&\frac{\pi c^2\abs\lambda^4}2\left[1+ \cos\left(\phi-\psi\right)\right]^2\delta(\mu) \\ {}&+& \frac{c^2\abs\lambda^2\varkappa^2} {2(\mu^2+\varkappa^2)}\left[1+ \cos\left(\phi-\psi\right)\right],
\end{eqnarray}
\begin{equation}\label{SIb}
S_{I_-}(\mu)=2\pi c^2\abs\lambda^4\left[\cos\left(\phi-\psi\right)\right]^2\delta(\mu) + \frac{c^2\abs\lambda^2\varkappa^2} {\mu^2+\varkappa^2}.
\end{equation}
By taking $\cos\left(\phi-\psi\right)=\pm 1$, one of the two spectra $S_{I_j}(\mu)$ can be made to vanish, because there is no light from one of the two ports. With the choice $\cos (\phi-\psi)=0$, only the $\delta$-peak in $S_{I_-}(\mu)$ disappears.

Note that no squeezing is produced when both the mirrors in the MZI are fixed and only linear elements are inserted in the optical circuit.

\section{The Hudson-Parthasarathy equation for the mechanical oscillator}\label{Sect:HPeq}

We consider now the case of a micro-mirror mounted on a vibrating structure and directly illuminated by a laser, so that it is subjected to a radiation pressure force. The mirror is only allowed to move perpendicularly to its plane. Being a vibrating mesoscopic body, also thermal effects on its motion have to be considered.

If we consider a well collimated laser beam and a perfect mirror, it is possible to represent the light by a single ray impinging on the mirror and reflected according to the laws of the geometrical optics. Then, the scattering part of the HP-equation  can describe this effect, as it was already introduced in \cite{Bar16,JG15}; other applications of the scattering interaction are given in \cite{BarP02,BarGQP13,GJN12,JG15}.
As usual, also the thermal effects can be modelled by the HP-equation and, indeed, QSC was initially introduced to treat quantum noise \cite{HudP84,Parthas92}; the specific case of a damped mechanical oscillator was formalized in \cite{BarV15,Bar16}.

HP-equation and quantum Langevin equations are based on QSC \cite{Parthas92}, an It\^o type-calculus involving the integral form \eqref{fdensity} of the  quantum fields. By using these fields a theory of quantum stochastic differential equations has been developed. In handling the ``stochastic differentials'' a ``promemoria'' is given by the It\^o table:
\begin{subequations}\label{Itotab}
\begin{eqnarray}
\rmd A_k(t)\rmd A_l^\dagger(t)&=&\delta_{kl}\rmd t,
\\  \rmd A_i(t) \rmd \Lambda^A_{kl}(t)&=& \delta_{ik} \rmd A_l(t),
\\
\rmd \Lambda^A_{kl}(t)\rmd A_i^\dagger(t)&=&\delta_{li}\rmd A_k^\dagger(t),
\\ \rmd \Lambda^A_{kl}(t)\rmd \Lambda^A_{ij}(t)&=&\delta_{li}\rmd \Lambda^A_{kj}(t),
\end{eqnarray}
\end{subequations}
all the other products vanish. This table has the same role of the heuristic rule $(\rmd W(t))^2=\rmd t$ in classical It\^o stochastic calculus.

\subsection{The equations of motion}\label{Sect:motion}
Thus, we consider the general HP-equation for a system on a Hilbert space $\Hcal$ interacting with two bosonic fields on the Fock space $\Gamma$, by pure scattering with the first one, and by pure absorption/emission with the second one:
\begin{eqnarray}\nonumber
\rmd U(t)&=&\biggl\{-\rmi H_\rmm\rmd t +  (S-\openone)\rmd \Lambda_{11}^B(t) \\ {} &+&
R\rmd A_{3}^\dagger(t) -R^\dagger\rmd A_{3}(t)-\frac 1 {2 }\,R^\dagger R \rmd t \biggr\}U(t),\label{HPeq}
\end{eqnarray}
with the initial condition $U(0) =\openone$. Here $H_\rmm$, $S$, $R$, are system operators; $H_\rmm$ is self-adjoint and $S$ is unitary, so that it can be written as $S=\rme^{\rmi \hat s}$ with $\hat s$ selfadjoint.

The solution $U(t)$ is a time-dependent unitary operator on $\Hcal\otimes \Gamma$ giving the global fields/system evolution in the interaction picture with respect to the free evolution of the fields. This latter is modelled by a time shift, while the free evolution of the system and the fields/system interaction are determined by the system operators $H_\rmm$, $S$, and $R$. The corresponding global Hamiltonian has been characterized in \cite{Greg01}. The unusual aspect of the evolution equation \eqref{HPeq} is due to the use of an It\^o type calculus and to the underlying rules \eqref{Itotab}. In writing \eqref{HPeq} we have taken $\hbar=1$. To have an idea of the system/field interaction it is useful to write at least the formal expression of the total Hamiltonian in the interaction picture, as done in \cite{ZolG97,Bar86} for the case without the scattering part. By using the rules summarized by \eqref{Itotab} one can check that the solution of \eqref{HPeq} can be written as the time ordered exponential
\begin{eqnarray*}
U(t)=\stackrel{\leftarrow}{T}\exp\int_0^t\Bigl\{&-&\rmi H_\rmm\rmd s \\ {} +
R\rmd A_{3}^\dagger(s) &-&R^\dagger\rmd A_{3}(s) + \rmi \hat s \rmd \Lambda_{11}^B(s)\Bigr\}.
\end{eqnarray*}
By using the field densities as in Sect.\ \ref{sec:MZI}, the formal total Hamiltonian, in the interaction picture, is
\begin{equation}\label{Htot}
H_{\rm tot}(t)=H_\rmm+\rmi R a_3^\dagger(t) -\rmi R^\dagger a_3(t)-\hat s b_1^\dagger(t)b_1(t);
\end{equation}
here we can recognize the system Hamiltonian,  the absorption/emission of phonons, and the interaction of the system with the incoming flow of photons.

Now we are going to select the system operators, as well the thermal field state, on the basis of phenomenological motivations. To develop this point we need
the evolution of a generic system operator $X$ in the Heisenberg description, which is given by $X(t)=U(t)^\dagger XU(t)$. By differentiating this product according to the rules of QSC \eqref{Itotab}, the HP-equation \eqref{HPeq} implies the following \emph{quantum Langevin equations} for $X(t)$:
\begin{eqnarray}\nonumber
\rmd X(t) &=& \rmi [H_\rmm(t),X(t)]\rmd t
\\ \nonumber {}&-&\frac 1 2  \bigl( R(t)^\dagger[R(t),X(t)] +
[X(t),R(t)^\dagger]R(t)\bigr) \rmd t
\\ \nonumber{}&+& [X(t),R(t)] \rmd A_3^\dagger(t)
-[X(t),R(t)^\dagger]\rmd A_3(t)
\\ {}&+& \biggl(S(t)^\dagger X(t)S(t)- X(t)\biggr)
\rmd \Lambda_{11}^B(t).\label{jdot}
\end{eqnarray}

In our case, the mechanical system is characterized by the position and momentum operators $q,\,p$, of which we take the dimensionless version: $[q,p]=\rmi$. Therefore, $\Hcal=L^2(\Rbb)$ and the interactions between the fields and the mirror depend on its position and momentum. Anyway, in the considered approximations, the dynamics of the system does not change the instant of the interaction between one boson and the system. Therefore, we are modelling an experiment where the system displacement has no appreciable effect on the optical paths.

The oscillating micromirror is a mesoscopic object, which is treated by an ``effective quantization'' of its position and momentum \cite[Sect.\ 2.2]{BM16}; when the pondemorotive interaction is not present it is treated in the linear approximation. Thus, the mechanical Hamiltonian $H_\rmm$ is taken to be a quadratic expression of $q$ and $p$. Then, the terms with the field operators $A_3$ must describe absorption/emission of phonons and must contribute to the damping of the oscillator; as the intrinsic motion is taken in the linear approximation, the operator $R$ is taken linear in $q,\,p$.

Finally, we have the ponderomotive interaction. The term with the unitary system operator $S$ must generate a force proportional to the rate of photon arrivals: $v\rmd  \Lambda_{11}^B(t)$, $v\in \Rbb$. The sign of $v$ depends on how the reference axis for the position is taken. With reference to Figure \ref{MZI} the axis along which the oscillator moves is vertical: if the positive direction is "up", $v$ is negative, if the positive direction is "down", $v$ is positive.

To fix the various operators, our first requirement is that the equations of motion for the means of position and momentum must be the classical ones for an oscillator. This means to ask that the quantum Langevin equations \eqref{jdot}, particularized to $q,\,p$, must take the form
\begin{subequations}\label{eq:qp}
\begin{eqnarray}
\rmd q(t)&=&\Om p(t) \rmd t +\rmd \hat W_q(t),
\\
\rmd p(t)&=&-\left(\Om  q(t) +\mgam  p(t)\right) \rmd t+v \rmd \Lambda_{11}^B(t)\nonumber
\\ &{}&\qquad\qquad\ {}+ \rmd \hat W_p(t).
\end{eqnarray}
\end{subequations}
Here, the elastic and damping forces and the radiation pressure force appear only in the equation for the momentum, as in the classical case. Then, the quantum noises $\hat W_q(t)$, $\hat W_p(t)$ must have vanishing means, in order to have the classical equations of motion for the means. The parameter $\Om>0$ is the bare frequency of the oscillator and $\mgam>0$ is the damping constant; moreover, we consider only the underdamped case: $\Omq>\mgamq/4$. The frequency $\Om$ appears also in the first equation, just for dimensional reasons; apart from the noise term, the first equation says that $p$ is proportional to the velocity. The idea of asking that the forces should appear only in the equation for the momentum is used also in \cite{Vitali12,BM16}, in the context of cavity optomechanics. The difference is the presence of noise in the first of \eqref{eq:qp}; this is due to the fact that our Langevin equations are not coming from some further approximation, but are an exact consequence of a unitary dynamics and that an It\^o-type calculus is used.

To obtain equations \eqref{eq:qp} from \eqref{HPeq}, \eqref{jdot}, it easy to check that we must have
\begin{subequations}\label{HRK}
\begin{eqnarray}\label{Hrmm}
&H_\rmm=H_0+ H_1, \qquad H_0=\frac{\Om}2\left(p^2+q^2\right),
\\ &H_1=\frac {\mgam }  4\left\{q,p\right\}, \qquad R= \alpha q +\beta p,\label{H1R}
\\
&\alpha\geq 0, \quad \beta\in \Cbb, \qquad 0<\alpha\IM \beta=\frac \mgam 2< \Om, \label{constants}
\\
&S=\rme^{\rmi \hat s}=\rme^{\rmi v q +\rmi \phi}, \qquad v\in\Rbb, \quad\phi\in[0,2\pi).\label{Sop}
\end{eqnarray}
\end{subequations}
We can always take $\alpha$ real and positive, because a phase can be included in the definition of $A_3$; $H_1$ represents a modification of the oscillator Hamiltonian due to the interaction with the phonon bath. The phase $\phi$ in the scattering operator $S$ will affect only the evolution of the fields and represents a phase shift due to the reflection; the parameter $v$ is a pure number and it depends on the incidence angle of the light and on the frequency of the laser. Some freedom remains in the operator $R$; it will be fixed by asking ``energy equipartition'' in the reduced equilibrium state of the oscillator, see Sect.\ \ref{Sect:physinter}.  When unbounded  operators are involved, as in \eqref{Hrmm}, some restrictions are needed in order to control the domains; then, existence, uniqueness, unitarity of the solution of \eqref{HPeq} can be proved \cite{FagW03}.

Finally, the quantum noises $\hat W_q(t)$ and $\hat W_p(t)$ come out from the terms in the third line of Eq.\ \eqref{jdot} and are given by
\begin{subequations}\label{C_qp}
\begin{eqnarray}
\hat W_q(t)&=& \rmi \beta A_3^\dagger(t) -\rmi \overline{\beta}\, A_3(t) ,
\\
\hat W_p(t)&=&\rmi \alpha \left( A_3(t)-A_3^\dagger(t)\right).
\end{eqnarray}
\end{subequations}
By construction, due to the unitarity of $U(t)$, the commutation relations for the system operators are preserved; also a direct verification is possible by showing that the quantum stochastic differential of $[q(t),p(t)]$ vanishes due to the commutation rules satisfied by the noises \eqref{C_qp}, given in \cite[(21)]{Bar16}. Moreover, our choice of the field state, given in Sect.\ \ref{sec:tfst}, will be such that the mean values of the noises $\hat W_q(t)$, $\hat W_p(t)$ are vanishing; then, the evolution equations for the mean values of $q$ and $p$ coming from \eqref{eq:qp} are exactly the classical equations for an underdamped oscillator.

The results on the mechanical Hamiltonian and the interaction oscillator-phonons have been obtained in \cite{BarV15} starting from symmetry requirements, while the scattering interaction with photons has been introduced in \cite{Bar16}. Here we have shown how to get the system/fields evolution by simplified arguments.

We can observe that
the motion of the mirror changes the optical path, and, similarly, the velocity of the mirror changes the frequency of the scattered photons (momentum conservation). These effects are not taken into account, as we expect them to be small. The mirror vibrations could also increase the dispersion of the $C_1$-beam; this effect could be corrected by lenses. Moreover, any change in the optical path in the lower arm of the MZI has to be compensated in the upper one. Another point is that we have considered a single vibrational mode of the mechanical oscillator, while physical oscillators could have more modes; this problem is present also in cavity optomechanics \cite[Sect.\ 2.2]{BM16}. Usually, it is assumed that a single mode is involved in the response to the optical force, but also the multimode case has been studied \cite{Vitali12}. As our presentation is aimed at giving a proof-of-principle of the production of non-classical light by pure scattering, we have considered the case of a response practically restricted to a single mechanical mode, while the theory could be generalized also to the multimode case.

\subsection{The mechanical mode operator and the solution of the Langevin equations}

The quadratic Hamiltonian $H_\rmm$ \eqref{Hrmm} can be diagonalized by introducing a suitable mode operator.
Firstly, we introduce the damped frequency $\om$ and the phase factor $\tau$ by
\begin{subequations}\label{om}
\begin{eqnarray}
\om&=&\sqrt{\Omq-\frac{\mgamq }4},
\\ \tau&=&\frac{\om}{\Om}-\frac{\rmi}{2} \frac{\mgam }{\Om}= -\frac\rmi\Om\left(\frac\mgam 2 +\rmi \om\right).
\end{eqnarray}
\end{subequations}
Then, we define the \emph{mode operator} for the mechanical oscillator
\begin{eqnarray}\nonumber
a_\rmm   &=& \sqrt{\frac{\Om}{2 \om}} \left( q +  {\rmi \tau} p\right)\\ {}&=& \frac{1}{\sqrt{2 \om\Om}} \left(\Om  q +  \frac \mgam 2\, p +\rmi \om  p\right),\label{aqptau}
\end{eqnarray}
satisfying the commutation rules $[a_\rmm,a_\rmm^\dagger]=1$; the inverse transformation turns out to be
\begin{equation}\label{atoqp}
q=\sqrt{\frac\Om{2\om}}\left(\overline \tau \, a_\rmm  +  \tau a_\rmm ^\dagger\right), \qquad p=\rmi\sqrt{\frac{ \Om}{2\om}}\left(a_\rmm ^\dagger - a_\rmm \right).
\end{equation}
Let us stress that the connection between the mode operator and the couple position-momentum is not the usual one, but the phase $\tau$ appears.
By these definitions, the mechanical Hamiltonian \eqref{Hrmm} can be rewritten in the form
\begin{equation}\label{Hm}
H_\rmm=H_0+H_1=\om \left(a_\rmm ^\dagger a_\rmm  +\frac 1 2 \right).
\end{equation}

Directly from \eqref{jdot} and \eqref{Hm}, or from \eqref{aqptau} and \eqref{eq:qp}, one gets the quantum Langevin equation for $a_\rmm$:
\begin{eqnarray}\nonumber
\rmd a_\rmm (t)&=&-\left(\rmi \om+\frac \mgam  2\right) a_\rmm (t)\rmd t \\ {}&+& \rmd \hat W_{a_\rmm}(t)+\rmi \tau  v\sqrt{\frac \Om{2\om}}\,\rmd \Lambda_{11}^B(t),\label{eq:a}
\\ \label{anoise}
\hat W_{a_\rmm}(t)&=& \sqrt{\frac\Om{2\om} } \left( \hat W_q(t)+\rmi \tau \hat W_p(t)\right).
\end{eqnarray}
The linearity of such equation allows for an explicit solution
\begin{eqnarray}\nonumber
a_\rmm (t)&=&\rme^{-\left(\rmi \om +\frac \mgam  2 \right)t}a_\rmm +  \int_0^t\rme^{-\left(\rmi \om +\frac \mgam  2 \right)\left(t-s\right)}\rmd \hat W_{a_\rmm}(s)
\\ {}&+&\rmi \tau v\sqrt{\frac \Om{2\om}}\int_0^t\rme^{-\left(\rmi \om +\frac \mgam  2 \right)\left(t-s\right)} \rmd \Lambda_{11}^B(s),
\label{a(t)}\end{eqnarray}
leading to the position and momentum Heisenberg operators
\begin{eqnarray}\nonumber
&&q(t) =\rme^{-\mgam  t/2} \left(q\cos \om t +\frac {\mgam  q+2\Om p}{2\om} \,\sin\om t\right)
\\ \nonumber&&{}+ \sqrt{\frac\Om{2\om} } \int_0^t\rme^{-\frac \mgam  2 \left(t-s\right)}\left[\rme^{-\rmi \om \left(t-s\right)}\overline{\tau}\,\rmd \hat W_{a_\rmm}(s)+\text{h.c.}\right]
\\  &&{}+
\frac{\Om v}\om\int_0^t \rme^{-\frac\mgam 2\left(t-s\right)}\sin \om\left(t-s\right)\,\rmd \Lambda_{11}^B(s),
\label{q(t)}
\end{eqnarray}
\begin{eqnarray}\nonumber
&&p(t) =\rme^{-\mgam  t/2} \left(p\cos \om t-\frac {2\Om q+\mgam  p}{2\om}\,\sin\om t\right)
\\ \nonumber &&{}- \sqrt{\frac\Om{2\om} } \int_0^t\rme^{-\frac \mgam  2 \left(t-s\right)}\Bigl[\rmi\rme^{-\rmi \om \left(t-s\right)} \rmd \hat W_{a_\rmm}(s)+\text{h.c.}\Bigr]
\\ \nonumber &&\qquad {}+
v\int_0^t \rme^{-\frac\mgam 2\left(t-s\right)}\biggl(\cos\om\left(t-s\right)
\\ && \qquad \qquad {}-\frac{\mgam}{2\om}\,\sin \om\left(t-s\right)\biggr)\rmd \Lambda_{11}^B(s).\label{p(t)}
\end{eqnarray}

\subsection{The state of the thermal field}\label{sec:tfst}

To describe a general thermal bath, we take as field state a suitable mixture of coherent states, as in \cite{BarV15,Bar16}.
Let $u$ be a stationary Gaussian complex random process with
\begin{subequations}\label{momu}
\begin{eqnarray}
&&\Ebb[u(t)]=0, \qquad \Ebb[u(t)\,u(s)]=0,
\\
&&\Ebb[\overline{u(t)}\,u(s)]=F(t-s),
\\
&&F(t) =\frac 1 {2\pi}\int_{-\infty}^{+\infty}\rme^{\rmi\nu t} N(\nu)\,\rmd \nu, \\
&&N(\nu)\geq 0, \qquad  N(\nu)\in L^1(\Rbb).
\end{eqnarray}
Thanks to stationarity, the function $F(t)$ is positive definite, so
that, according to Bochner's theorem \cite[Theor.\ IX.9]{RS80}, its Fourier transform
\begin{equation}   \label{Nu}
N(\nu)=\int_{-\infty}^{+\infty}\rme^{-\rmi\nu t}F(t)\,\rmd t
\end{equation}
\end{subequations}
is a positive function, which we assume to be absolutely integrable,
thus implying a finite power spectral density for the process. The quantity $N(\nu)$ will play the role of noise spectral density in the dynamics of the oscillator; it could be the Bose-Einstein distribution or any other temperature dependent function.

We take the state of the thermal field $A_3(t)$ to be  the mixture of coherent states
\begin{equation}\label{uTst}
\rho_{\rm th}^T =\Ebb\left[|e_{3}(u_T)\rangle\langle e_{3}(u_T)|\right], \qquad u_T(t):= \ind_{[0,T]}(t) u(t).
\end{equation}
The quantity $u$ is a complex stochastic process with locally square integrable trajectories and $\Ebb$ denotes the expectation with respect to the probability law of the process $u$. In the argument of a coherent vector only square integrable functions are allowed, while the trajectories of the process $u$ are only locally square integrable, i.e. $\int_{t_0}^{t_1} \abs{u(t)}^2\rmd t<+\infty$ for all time intervals $(t_0,t_1)$ of finite length. So, we have introduced the cutoff $T$,  representing a large time, which we will let tend to infinity in the final formulae describing the quantities of direct physical interest, as for the state of the optical fields. As explained in \cite[Sect.\ 3.2.1]{BarV15} this is a field analog of the regular $P$-representation of the case of discrete modes \cite{GarZ00}.

\subsection{The equilibrium state of the quantum oscillator}\label{sec:equil}
\label{Sect:physinter}
When the field state $\rho_{\rm field}^T=\rho_{\rm em}^T\otimes \rho_{\rm th}^T$ is the vacuum state or, more generally, a coherent vector, the reduced system state $\rho_\rmm(t)$ satisfies a Markovian master equation \cite{Parthas92,Bar06} with a Lindblad type generator \cite{Lin76,Lin76b}. If a more general state is taken for the field state, non-Markov effects enter into play and a simple closed evolution equation for the reduced dynamics could even not exist \cite{Bar06,GJN12}. Indeed, this is the case for the thermal state introduced in Sect.\ \ref{sec:tfst}.
While the reduced state for the mechanical oscillator does not satisfy a simple closed master equation,
in principle all the properties of the mechanical oscillator
can be computed (without relying on a master equation), because we have the explicit solutions of the Langevin equations and the quantum correlations of the fields introduced in Sects.\ \ref{Sect:fs}, \ref{sec:tfst}.

By using system Weyl operators in the Heisenberg picture one could prove the existence and characterize the
reduced  equilibrium state of the quantum oscillator
\[
\rho_\rmm^{\rm eq}=\lim_{T\to +\infty}\Tr_\Gamma \left\{U(T) \left(\rho
_\rmm\otimes \rho_{\rm field}^T
  \right)U(T)^\dagger\right\},
\]
where $\Tr_\Gamma$ denotes the partial trace over the fields.
However, in this work we need only the first and second moments of position and momentum at equilibrium. By taking the quantum expectation of \eqref{q(t)} and \eqref{p(t)} and the limit for large times, we obtain
\begin{equation}\label{Epq}
\langle p\rangle_{\rm eq}=0, \qquad \langle q\rangle_{\rm eq}=\frac{v\eta \abs\lambda^2}\Om=:q_\infty.
\end{equation}
The same procedure can be applied to $q(t)^2$ and $p(t)^2$. We can introduce now the last requirement to fix completely the system operator $R$.
\begin{assumption}
We ask to have \emph{energy equipartition }  in the mean for the fluctuation contributions, i.e.\ we require
\[
\langle q^2\rangle_{\rm eq}-q_\infty^2= \langle p^2\rangle_{\rm eq};
\]
moreover, we ask this equipartition to hold for any temperature, i.e.\ independently from the choice of $N(\nu)$ in the definition of the state of the thermal field, given in Sect.\ \ref{sec:tfst}.
\end{assumption}

The computations of the second moments are long; the results are reported in Appendix \ref{qpfl}.  The Assumption above can be applied to \eqref{momq2}, \eqref{momp2}, and this fixes the residual freedom in the definition of the operator $R$ given in \eqref{H1R}; the final result is
\begin{equation}\label{Rfinal}\begin{split}
\alpha=\sqrt{\frac{\mgam\Om}{2\om}}, &\qquad \beta=\rmi \tau \sqrt{\frac{\mgam\Om}{2\om}}\\ \Rightarrow \quad &R=\sqrt{\mgam}\,a_\rmm.
\end{split}\end{equation}
So, all the system operators in the HP-equation \eqref{HPeq} are now fixed and the same holds for the quantum noises \eqref{C_qp} and \eqref{anoise}. As reported in Appendix \ref{qpfl}, the final expressions of the second moments turns out  to be
\begin{equation}
\langle q^2\rangle_{\rm eq}-q_\infty^2=\langle p^2\rangle_{\rm eq}=\frac\Om\om \left(N_{\rm eff}+\frac1 2 \right)+\frac{\eta \abs\lambda^2 v^2}{2\mgam},\label{q2p2}
\end{equation}
\begin{equation}
\langle\left\{ q,p\right\}\rangle_{\rm eq}= -\frac\mgam{\om} \left(N_{\rm eff}+\frac1 2 \right), \label{q,p}
\end{equation}
\begin{equation}
N_{\rm eff}:= \frac \mgam{2\pi}\int_\Rbb \frac{N(\nu)}{\frac{\mgamq}4+\left(\om-\nu\right)^2}\,\rmd \nu. \label{Neff}
\end{equation}

As we have already observed, the solution $U(t)$ of the HP-equation is the unitary evolution operator in the interaction picture with respect to the free-field dynamics. To physically understand this dynamics and to visualize the system/field interaction, it is useful  to discuss the formal Hamiltonian \eqref{Htot}. By \eqref{HRK}, \eqref{aqptau}, \eqref{Hm}, \eqref{Rfinal}, we get
\begin{eqnarray*}
H_{\rm tot}(t)&=&
\om \left(a_\rmm ^\dagger a_\rmm  +\frac 1 2 \right)+\rmi \sqrt{\mgam}\,a_\rmm a_3^\dagger(t)\\ {}&-&\rmi \sqrt{\mgam}\,a_\rmm ^\dagger a_3(t)
-\left(vq+\phi\right) b_1^\dagger(t)b_1(t).
\end{eqnarray*}
The first three terms represent the system Hamiltonian and the system/phonon interaction;  the appearance is that of an optical mode with absorption/emission interaction with a Bose field. However, the mechanical mode operator is connected to position and momentum in an unusual way \eqref{aqptau}; no rotating wave approximation is involved and the effects on the equations of motion \eqref{eq:qp} are the noise terms
$\rmd \hat W_q(t)$, $\rmd \hat W_p(t)$ and the presence of the damping term in the equation for the momentum. The final term is the strict analog of the interaction used in all the works in cavity optomechanics and it gives the radiation pressure force in the equation for momentum. Let us recall that the peculiar feature of the HP-equation is that the motion is given in the singular limit of vanishing interaction time and that the resulting  evolution equation is the HP-equation \eqref{HPeq}.

As already observed at the beginning of Sect.\ \ref{sec:equil}, in the case of a generic noise spectrum, it does not exist a closed master equation for the reduced state of the mechanical oscillator; this is not a problem because all the following computations will be based on the solution \eqref{q(t)} of the Heisenberg equations of motion \eqref{eq:qp}. Only in the limiting case of constant phonon spectrum, $N(\nu)\to N_{\rm eff}$ a Markovian master equation is obtained (without involving new approximations), given in \cite[p.\ 326]{Bar16}. The Liouville operator can be written in Lindblad form \cite{Lin76,Lin76b}. It contains a term with the structure of the generator of a Poisson semigroup, due to the contribution of the kicks of the photons, while the other terms have the structure of an optical master equation. Once again, this is not due to some rotating wave approximation, but to the unusual definition of the mechanical mode operator. The Heisenberg equations of motion \eqref{eq:qp} do not depend on the noise spectrum and they describe a mechanical Brownian oscillator.

\subsection{Input-output relations and the scattering operator}\label{sec:IO}

Also the fields in the Heisenberg picture can be introduced \cite{GarC85,Bar86}; these are known as \emph{output fields} and they have been defined in \eqref{C=out}.
The outputs fields $C_{1}(t)$, $C_{1}^\dagger(t)$, $\Lambda_{11}^C(t)$ represent the fields after the interaction with the system, while $B_{1}(t)$, $B_{1}^\dagger(t)$, $\Lambda_{11}^B(t)$ are the fields before the interaction and, so, they are called \emph{input fields}. By the properties of $U(t)$ we get, $\forall T\geq t$,
\begin{eqnarray*}
U(T)^\dagger B_1(t)U(T)&=&U(t)^\dagger B_1(t)U(t)\equiv C_1(t), \\ U(T)^\dagger \Lambda_{11}^B(t)U(T)&=&U(t)^\dagger \Lambda_{11}^B(t)U(t)\equiv \Lambda_{11}^C(t).
\end{eqnarray*}
This implies that the output fields satisfy the same CCRs as the input fields, as anticipated in Remark \ref{rem:Tt}.
Self-adjoint combinations of the output fields commuting for different times represent field observables which can be measured with continuity in time and this is the key ingredient for a quantum theory of measurements in continuous time \cite{Bar06,BarCS11,Bar86,BarG13}. Output fields were introduced independently in quantum optics in connections with quantum Langevin equations and are often used \cite{GarC85,ZolG97,GarZ00,WisM10,GarPC87}.

By differentiating the products defining the output fields \eqref{C=out} and using \eqref{HPeq} and \eqref{Itotab}, we get the input/output relations \cite{Bar06}
\begin{eqnarray}\label{Beminout}
\rmd  C_{1}(t)&=&S(t)\rmd B_{1}(t)=\rme^{\rmi vq(t)+\rmi\phi}\rmd B_{1}(t),
\\ \label{2Ls}
\rmd \Lambda^{C}_{11}(t)&=&S(t)^\dagger S(t) \rmd \Lambda_{11}^B(t)
=\rmd \Lambda_{11}^B(t).
\end{eqnarray}
Note that the number operator for the photons is not changed by the interaction with the mirror.

By using the field densities we can write Eq.\ \eqref{Beminout} as
\begin{equation}\label{keyeq}
c_1(t)=\rme^{\rmi vq(t)+\rmi\phi}b_1(t).
\end{equation}
This is the key result which allows for the computation of the various observed quantities by inserting it into \eqref{meanN3}--\eqref{Sigmas}. As already stressed, by \eqref{2Ls} no trace of the interaction is contained in the number process for the field $C_1$; so, only after some kind of interference with a reference field, the ``quantum phase'' introduced by the interaction can be detected. Indeed, in our scheme the fields $C_1(t)$ and $C_2(t)$ are made to interfere at the second beam splitter and they are the $D$-fields at the output ports  which are monitored.

Another important point is that the motion of the oscillator will depend on the intensity of the incoming light; so, the term $\rme^{\rmi vq(t)}$ should introduce an intensity dependent phase in the field. In quantum optics, it is expected  that this situation could squeeze the light \cite[Sect.\ 4.5]{BM16}. Indeed, we shall see typical effects of squeezing in the various detection schemes discussed in Section \ref{Sect:det}. In Ref.\ \cite{BM16} the case of cavity optomechanics is considered and the term \emph{ponderomotive squeezing} is introduced; this terminology is well suited also in our case, as in both cases the radiation pressure interaction is involved.

To apply \eqref{keyeq} in explicit computation, we need to elaborate the scattering operator $\rme^{\rmi v q(t)}$.
Let us define the functions
\begin{eqnarray}\label{ell(t)}
\ell_t(s)&:=&l(t-s)\ind_{(0,t)}(s),\\ \nonumber l(t)&:=&-\rmi v\tau \sqrt{\frac{\Om \mgam  }{2\om}}\,\rme^{-\frac \mgam  2\abs t}\rme^{\rmi \om t},
\\ \label{Vts}
V_t(s)&:=&\exp\left\{2\rmi  h(t-s)1_{(0,t)}(s)\right\},\\ h(t)&:=& \frac{v^2\Om}{2\om}\,  \rme^{-\frac\mgam2\abs t}\sin \om t.\label{def:h(t)}
\end{eqnarray}
We consider $V_t$ as the unitary operator defined by
\begin{equation}\label{V(t)}
\bigl(V_tw\bigr)(s)=V_t(s)w(s),\qquad \forall w\in L^2(\Rbb).
\end{equation}
Then, \eqref{q(t)} can be written as
\begin{eqnarray*}
\rmi v q(t) &=&\rmi v\rme^{-\mgam  t/2} \left(\rme^{-\rmi\om t}\overline \tau\,a_{\rmm}+\rme^{\rmi\om t} \tau\,a_{\rmm}^\dagger\right)
\\ {}&+&\int_0^t\,l(t-s)\rmd A_{3}^\dagger (s) -\int_0^t\,\overline{l(t-s)}\,\rmd A_{3} (s)
\\ {}&+&2\rmi
\int_0^t h(t-s)\,\rmd \Lambda_{11}^B(s).
\end{eqnarray*}

Take now the scattering operator \eqref{Sop}; in the Heisenberg picture we have
\begin{subequations}\label{SWW}
\begin{eqnarray}
S(t)&=&U(t)^\dagger S U(t)=\rme^{\rmi v q(t)+\rmi \phi}, \\ \rme^{\rmi v q(t)}&=&S_0(t)W_3(\ell_t)W_1(V_t);
\end{eqnarray}
\end{subequations}
in this decomposition a system operator and two Weyl operators appear:
\begin{eqnarray}\nonumber
S_0(t)&=&\exp\biggl\{\rmi v \rme^{-\mgam  t/2} \biggl(q\cos \om t\\ {} &+&\frac {\mgam  q+2\Om p}{2\om} \,\sin\om t\biggr)\biggr\}\overset{t\to +\infty}{\longrightarrow} 1,\label{S0to1}
\\ \label{Wem}
W_1(V_t)&=&\exp\left\{2\rmi
\int_0^t h(t-s)\,\rmd \Lambda_{11}^B(s)\right\},
\\ \label{Wth}
W_3(\ell_t)&=&\exp\left\{\int_0^{+\infty} \ell_t(s)\rmd A_{3}^\dagger(s) -\text{h.c.}\right\}.
\end{eqnarray}
The operator $W_1(V_t)$ \eqref{Wem} is a Weyl operator acting only on the electromagnetic component and characterized by the unitary operator $V_t$ \eqref{V(t)}, and $W_3(\ell_t)$ \eqref{Wth} is a displacement operator with function $\ell_t$ \eqref{ell(t)} acting on the thermal component.

From this decomposition, some useful properties of the scattering operator follow.
\begin{proposition}\label{proposition} The following identities hold:
\begin{equation}\label{commutS_0}
V_s(t)\rme^{\rmi \IM \langle \ell_s|\ell_t\rangle}S_0(t)S_0(s)=V_t(s)\rme^{-\rmi \IM \langle \ell_s|\ell_t\rangle}S_0(s)S_0(t),
\end{equation}
\begin{equation}\label{Vqq}
V_t(s)\rme^{\rmi v q(s)}\rme^{\rmi v q(t)}=V_s(t)\rme^{\rmi v q(t)}\rme^{\rmi v q(s)},
\end{equation}
\begin{equation}\label{Vqqq}
V_t(s)\rme^{-\rmi v q(t)}\rme^{\rmi v q(s)}\rme^{\rmi v q(t)}=V_s(t)\rme^{\rmi v q(s)},
\end{equation}
\begin{equation}\label{[,]}
b_1(s)\rme^{\rmi v q(t)}=\rme^{\rmi v q(t)}V_t(s)b_1(s).
\end{equation}
\end{proposition}
The proof of this Proposition is given in Appendix \ref{proof:prop}.

\begin{remark} \label{rem:fix} To get the case of a fixed mirror we can send the mass of the oscillator to infinity. One has to introduce the position and momentum operators in physical units, by which one sees that $v= g_0/\sqrt{\hbar m \Om}$, where $g_0$ is the radiation pressure constant in physical units. Letting $m\to+\infty$ we get $v=0$. Then, the
action of the evolution on the field \eqref{keyeq} reduces to $c_1(t)= \rme^{\rmi \phi}b_1(t)$, which is the case considered in Sect.\ \ref{noQ}. So, in this case the dynamics $U(t)$ plays the role of a Weyl operator, as it is in the upper arm, where a shift $\psi$ was introduced, due to the fixed mirror and a phase shifter. Indeed, any Weyl operator can be obtained by means of a suitable HP-equation and the whole optical circuit can be seen as a network where the output field from a node becomes the input field in another node. Indeed, such  networks are studied, for instance, in \cite{GJ09} and similar ideas are below the \emph{cascaded systems} in \cite[Sect.\ 19.2]{Car08}.
\end{remark}

\section{Detection of the output light}\label{Sect:det}
By considering the concrete model of micromirror/fields interaction via radiation pressure, giving rise to the connection \eqref{keyeq} of the field $c_1(t)$ with the scattering operator, we obtain the reduced spectra \eqref{Sigmas} and the mean rate \eqref{meanN3} in terms of the scattering operator \eqref{SWW}; this is done in Appendix \ref{Dmom_comp}.
Then, the moments of the scattering operator can be elaborated and computed (see Appendices \ref{sec:momWem}--\ref{mainmoments}); in this way we obtain the final formulae for the variances and the spectra.

\paragraph{Some useful notations.} To abbreviate the final expressions we need some definitions. We recall that we have already introduced $\ell(t)$ in \eqref{ell(t)},  $h(t)$ in \eqref{def:h(t)}, $N_{\rm eff}$ in \eqref{Neff}; we need also
\begin{eqnarray}\label{<3>}
\chi&:=&2\sqrt{\eta \left(1-\eta  \right)}, \quad \chi \in [0,1],
\\ \alpha&:=&\psi- \phi-\theta, \qquad K:=\frac{\Om v^2}{2\om}
\left(N_{\rm eff}+\frac 1 2\right),\label{def:K}
\\ \label{def:M}
M&:=&\eta  \abs{\lambda}^2 \int_0^{+\infty}\rmd s \left[1-\cos 2h(s)\right],
\\ \label{def:theta}
\theta&:=&\eta \abs\lambda^2\int_0^{+\infty}\rmd s \, \sin 2h(s),
\end{eqnarray}
\begin{eqnarray}\nonumber
g(t)&:=&\frac {v^2\Om}{2\om}\biggl[\rme^{-\frac\mgam 2 \abs{t}}\cos\om t \\ {}&+& \int_\Rbb \rmd \nu\, \frac{\mgam N(\nu)\cos\nu t}{\pi\left[ \left(\nu-\om\right)^2+\mgam^2/4\right]} \biggr].\label{def:g}
\end{eqnarray}

\begin{remark}[The mean rate of counts]\label{rem:nj} By the computations given in the appendices (Eqs.\ \eqref{nj:q} and \eqref{for_mean}), we obtain the expressions of the mean rates of counts at the two output ports:
\begin{equation}\label{n_jfinal}
n_j  = \frac{\abs\lambda^2}2\left[1
+(-1)^j  \chi \rme^{-(K+M)}\cos\alpha\right],
\end{equation}
from which we get
\begin{equation}\label{n+-}
n_1+n_2=\abs\lambda^2, \qquad n_2-n_1=\abs\lambda^2\chi\rme^{-(K+M)}\cos\alpha.
\end{equation}
\end{remark}

\subsection{The spectrum  of the output currents}

The analytic expressions of the reduced spectra \eqref{Sigmas} are obtained in Appendix \ref{Dmom_comp}:
\begin{subequations}\label{redsp}
\begin{eqnarray}\nonumber
&&\Sigma_-^0(\mu)=2\abs\lambda^2\eta\rme^{-2(K+M)} \RE\int_0^{+\infty} \rmd s\, \rme^{\rmi\mu s}
\\ \nonumber{}&&\times\biggl[
\exp\biggl\{\eta \abs\lambda^2\int_0^{+\infty}\rmd u \left(\rme^{- 2\rmi h(u)}-1\right)
\\ {}&&\times\left( \rme^{2\rmi h(s+u)} -1\right)
+ g(s) +\rmi h(s)
\biggr\}-1+\text{c.c.}\biggr]\!,
\label{compSigma0}\end{eqnarray}
\begin{eqnarray}\nonumber
{}\!\!\!&&\Sigma_-^\psi(\mu)=2\abs\lambda^2\eta\rme^{-2(K+M)} \RE\int_0^{+\infty} \rmd s\, \rme^{\rmi\mu s}
\\ \nonumber
{}&&\!\!\times\!\biggl\{\rme^{-2\rmi \alpha} \biggl[\exp\biggl\{
\eta \abs\lambda^2\int_0^{+\infty}\rmd u \left(\rme^{2\rmi h(u)} -1\right)
\\ {} &&\!\!\times\!\left(\rme^{2\rmi h(s+u)} -1\right)-g(s)+\rmi h(s)\biggr\}-1\biggr]+\text{c.c.}\!\biggr\}\!,
\label{compSigmapsi}\end{eqnarray}
\begin{eqnarray}\nonumber
&&\Sigma_0(\mu)=2\rme^{-(K+M)}\eta^{3/2}\abs\lambda^2
\\ &&{}\!\times\RE\int_0^{+\infty} \rmd t\, \rme^{\rmi\mu t}\left[\rme^{-\rmi \alpha}\left(\rme^{2\rmi h(t)}-1\right)
+\text{c.c.} \right]\!,\label{Sigma+0}
\end{eqnarray}
\begin{equation}\label{+=0}
\Sigma_+(\mu)=0.
\end{equation}
\end{subequations}
We stress that the vanishing of the component $\Sigma_+(\mu)$ is due to the scattering structure of the interaction leaving $\Lambda^D_{11}(t)$ invariant, see \eqref{2Ls}. Let us recall that $\Sigma_-(\mu)=\Sigma_-^0(\mu)+\Sigma_-^\psi(\mu)$ and that \eqref{90b} holds. For $v=0$, which is equivalent to the case of a fixed mirror (see Remark \ref{rem:fix} and  Sect.\ \ref{noQ}),  we have $g(s)=h(s)=0$ and all the reduced spectra \eqref{redsp} vanish.

Then, by \eqref{+=0}, from \eqref{S+}--\eqref{Sj}, \eqref{n_jfinal}, \eqref{n+-}, we obtain the final expressions of the intensity spectra:
\begin{eqnarray}\nonumber
&&S_{I_j}(\mu) = 2\pi c^2 n_j^2\delta(\mu)
+\frac{c^2\varkappa^2\abs\lambda^2}{4\left(\mu^2 +\varkappa^2\right)}
\Bigl[2
+(-1)^j\sqrt{1-\eta  }\\ &&\times \left(  4\sqrt\eta \rme^{-(K+M)}\cos\alpha+\Sigma_0(\mu) \right)
+\left(1-\eta\right)\Sigma_-(\mu)\Bigr],
\label{SIjfinal}\end{eqnarray}
\begin{eqnarray}\nonumber
S_{I_-}(\mu) &=&2\pi c^2\abs\lambda^4\chi^2\rme^{-2\left(K+M\right)}\left(\cos\alpha\right)^2\delta(\mu)\\ {}&+&\frac{c^2\varkappa^2\abs\lambda^2}{\mu^2 +\varkappa^2}\left[1+\left(1-\eta\right)\Sigma_-(\mu)\right],
\label{S-final}\end{eqnarray}
\begin{equation}\label{S+final}
S_{I_+}(\mu) =2\pi c^2\abs\lambda^4\delta(\mu)+\frac{c^2\varkappa^2\abs\lambda^2}{\mu^2 +\varkappa^2} .
\end{equation}
Let us recall that the terms proportional to the Dirac deltas are the contribution of the constant part of the various currents, the factor $\frac{c^2\varkappa^2}{\mu^2 +\varkappa^2}$ is the square modulus of the Fourier transform of the detector response function, the constant terms inside the square brackets are the contributions of the shot noise.

\subsubsection{The reduced spectra}\label{sec:redspectra}

The reduced spectrum $\Sigma_{-}(\mu)$ can be obtained from the measurements of $S_{I_-}(\mu)$ and $S_{I_+}(\mu)$. In these two spectra the sharp peaks in $\mu=0$ can be individuated and subtracted. Then, the contribution of the detector response function is estimated from
\begin{equation*}
\frac{c^2\varkappa^2\abs\lambda^2}{\mu^2 +\varkappa^2 }=S_{I_+} (\mu) -2\pi c^2\abs\lambda^4\delta(\mu).
\end{equation*}
Finally, the reduced spectrum $\Sigma_{-}(\mu)$ is estimated from the measurement of the $I_-$-spectrum \eqref{S-final} by taking
\begin{eqnarray*}
\Sigma_-(\mu)&=&\frac {\varkappa^2+\mu^2}{c^2\varkappa^2 \abs{\lambda}^2}\,S_{I_-}(\mu)-1\\ {}&-&2\pi\abs\lambda^2\chi^2 \left(\cos\alpha\right)^2  \rme^{-2(K+M)}\delta(\mu).
\end{eqnarray*}

As discussed in Sect.\ \ref{sect:squeeze}, when the reduced spectrum is negative, the shot noise $n_1+n_2$ is reduced and this effect is due to the presence of squeezing in the light in field $C_1$. According to Eq.\ \eqref{cmurepr}, $1+\Sigma_-(\mu)$ is equal to the variance of the quadrature $Q_T(\mu;\psi)$ \eqref{Qmu}. Let us discuss here the sign of $\Sigma_-(0)$ in the general case. By defining
\begin{eqnarray}\nonumber
&Z&:=4\abs\lambda^2\eta\rme^{-2(K+M)}  \int_0^{+\infty} \rmd s \biggl[1
-\exp\biggl\{-g(s)+\rmi h(s) \\ {}&+&\eta \abs\lambda^2\int_0^{+\infty}\rmd u \left(\rme^{2\rmi h(u)} -1\right) \left(\rme^{2\rmi h(s+u)} -1\right)\biggr\}\biggr],
\label{defZ}\end{eqnarray}
we can write
\[
\Sigma_-(0)=\Sigma_-^0(0)-\RE\left(\rme^{-2\rmi \alpha} Z\right).
\]
By choosing  $\psi=\psi_0$ such that $2\alpha_0=\arg Z$, we get the minimum possible value of $\Sigma_-(0)$:
\begin{equation}
\Sigma_-(0)\big|_{\psi=\psi_0}=\Sigma_-^0(0)-\abs Z.
\end{equation}
As we shall see, for certain values of the parameters it becomes negative. We recall that $\Sigma_-^0(0)\geq 0$. By the fact that the functions $g(s)$ \eqref{def:g} grows with the temperature, by comparing \eqref{compSigma0} and \eqref{defZ}, we see that at high temperature this minimum value is positive.

By taking instead $\psi=\psi_1=\psi_0\pm \frac \pi 2$, we get the maximum value:
\begin{equation}
\Sigma_-(0)\big|_{\psi=\psi_1}=\Sigma_-^0(0)+\abs Z.
\end{equation}
By the two bounds \eqref{2bounds}, we have
\begin{eqnarray}\nonumber
1&\leq& \left(1+\Sigma_-(0)\big|_{\psi=\psi_0}\right)\left(1+\Sigma_-(0)\big|_{\psi=\psi_1}\right)
\\ \nonumber
{}\qquad &&=\left(1+\Sigma_-^0(0)\right)^2-\abs Z^2,
\\ \label{starting}
-1&\leq& \Sigma_-(0)\big|_{\psi=\psi_0}=\frac{\Sigma_-^0(0)^2-\abs Z^2}{\Sigma_-^0(0)+\abs Z}.
\end{eqnarray}

Aside from the value in $\mu=0$, we can say in general that the reduced spectra vanish for very large $\mu$:
\begin{equation}
\lim_{\mu\to \pm \infty}\Sigma_-(\mu)=0, \qquad \lim_{\mu\to \pm \infty}\Sigma_0(\mu)=0.
\end{equation}

\paragraph{The reduced spectrum $\Sigma_0(\mu)$.}
As done for $\Sigma_-(\mu)$, also the reduced spectrum $\Sigma_0(\mu)$ can be estimated from measurements of the intensity spectra \eqref{SIjfinal} and of the mean fluxes of photons \eqref{n_jfinal}.
From \eqref{def:M}, \eqref{def:theta}, \eqref{Sigma+0} we see that we have
\begin{equation}\label{Sigma00}
\Sigma_0(0)= 4\rme^{-(K+M)}\sqrt\eta\left(\theta \sin\alpha-M\cos \alpha\right).
\end{equation}

From \eqref{SIjfinal} we see also that, when $\Sigma_-(0)$ is negative, there is shot noise reduction in at least one of the intensity spectra, which means that the presence of non-classical light can be detected  in at least one of the two monitored beams.

\subsubsection{Weak interaction and strong laser}
The expressions \eqref{redsp}  of the reduced spectra are very involved, as they contain integrals of exponentials of functions and of other integrals\ldots So, in order to have some idea of their behaviour and to see if light squeezing is present, we need some approximation. In the following we shall study the case of weak interaction and strong laser and we shall also show  that in this extreme case a strong squeezing appears.

\begin{remark} [The approximation of flat noise spectrum]\label{rem:flat} From now on we use the following approximation: the spectral function $N(\nu)$ is
slowly varying in a neighbourhood of $\om$ of width $\mgam$. By \eqref{def:h(t)}, \eqref{Neff}, \eqref{def:g} this approximation gives
\begin{eqnarray}\nonumber
N_{\rm eff}&\simeq& N(\om), \\ \nonumber g(t)&+&\rmi h(t)\simeq\frac {v^2\Om}{2\om}\, \rme^{-\frac\mgam 2  \abs t }\\ {}&\times&\left[ \left(N(\om)+1\right)\rme^{\rmi \om t}+  N(\om)\rme^{-\rmi \om t}\right] .
\label{Nflat}\end{eqnarray}
\end{remark}

The radiation pressure interaction is certainly weak; moreover, in principle the interaction parameter $\abs v$ can be changed by changing the angle of incidence on the mirrors of the MZI.
\begin{remark}[Weak interaction]\label{wint} The parameter $v^2$ is small and the temperature is not too high; precisely, we assume to have
\begin{subequations}\label{vsmall}
\begin{eqnarray}
\left(N_{\rm eff}+\frac 12\right)\frac {v^2\Om}{2\om}&\ll& 1,
\\ \bigg(\frac{\eta\abs\lambda^2 v^4\Om}{4\mgam\om }\bigg)^2\ll \frac {v^2\Om}{2\om}&\ll& 1.
\end{eqnarray}
\end{subequations}
In particular, these conditions imply  $K+M\simeq 0$.
\end{remark}

By using these conditions it is sufficient to take only the first order terms in the formulae \eqref{redsp} for the reduced spectra; by the computations given in Appendix \ref{sec:approx}, we obtain the following  expressions
\begin{eqnarray}\nonumber
&&\Sigma_-(\mu)\simeq\frac{2\eta\abs\lambda^2v^2\Om^{\;3}}{\left[\frac\mgamq 4 + \left(\mu+\om\right)^2\right]\left[\frac\mgamq 4 + \left(\mu-\om\right)^2\right]}
\\ &&\ {}\times \left[\left(1 -\frac{\mu^2}\Omq\right) \sin 2\alpha
+E(\mu)
\left(1- \cos 2\alpha\right)\right],
\label{Sigma-approx}\end{eqnarray}
\begin{equation}
E(\mu):=\frac{\eta\abs\lambda^2 v^2}\Om+\frac{\mgam}{2\om}\left(2N(\om)+1\right)\left(1+\frac{\mu^2}\Omq\right),
\end{equation}
\begin{equation}\label{Sigma0approx}
\Sigma_0(\mu)\simeq
\frac {4\eta^{3/2}\abs\lambda^2v^2\Om\left(\Omq-\mu^2\right)\sin\alpha}  {\left[\frac\mgamq 4+\left(\mu+\om\right)^2 \right]\left[\frac\mgamq 4+\left(\mu-\om\right)^2\right]}.
\end{equation}

Let us note that, by series expansions of the exponentials in the formulae \eqref{redsp}, we would have obtained a representation of the spectra in terms of peaks centered in $\mu=\pm n\om$. Equations \eqref{Sigma-approx}, \eqref{Sigma0approx} give the relevant peaks under the assumption \eqref{vsmall}. Without the assumption of flat noise spectrum (Remark \ref{rem:flat}), also a smooth component would appear inside the peak structure of the optical spectrum.

\subsubsection{Production and detection of squeezed light}\label{sqdetect}

As discussed in Sect.\ \ref{sect:squeeze}, the squeezing of the light in the field $C_1$ is revealed by the negativity of $\Sigma_-(\mu)$ for some choice of the tunable phase $\psi$. Let us consider the case $\mu=0$  in \eqref{Sigma-approx};
as done  in Sect.\ \ref{sec:redspectra} for the general case, the minimum of the approximated expression of $\Sigma_-(0)$ is for $\psi=\psi_0$ such that
\begin{equation}\label{2alpha0}
\sin 2\alpha_0=-\frac1{\sqrt{E(0)^2+1}},\qquad \cos2\alpha_0=\frac {E(0)}{\sqrt{E(0)^2+1}}.
\end{equation}
With this choice we get
\begin{eqnarray}\nonumber
\Sigma_-(0)&&\big|_{\psi=\psi_0}=\Sigma_-^0(0)+\Sigma_-^{\psi_0}(0)\\ {}\simeq&&
\frac{2\eta \abs\lambda^2v^2}{\Om}\left(E(0) -\sqrt{E(0)^2+1}\right).
\label{min}\end{eqnarray}
This minimum is negative, but it can be far from the theoretical bound $-1$. Moreover, the whole spectrum \eqref{Sigma-approx} can be too small.
So, to overcome these drawbacks we ask to have a sufficiently strong laser, which however has to satisfy  \eqref{vsmall}. Let us note that the configuration represented by the second beam splitter, the two detectors, subtraction of the currents, very bright reference beam (i.e.\ $\eta$ small) is indeed the configuration of balanced homodyne detection.

\begin{remark}[Strong laser]\label{stronglaser} We assume the laser to be strong enough to satisfy
\begin{equation}\label{strlaser}
\frac{\eta\abs\lambda^2v^2}{\Om}\gg \frac{\mgam}{\om}\left(N(\om)+\frac 12 \right), \qquad E(0)^2\gg 1.
\end{equation}
\end{remark}
From a mathematical point of view the requirements \eqref{strlaser} and \eqref{vsmall}  are compatible and can be realized by keeping constant the product $L:= \eta\abs\lambda^2 v^{2(1+\epsilon)}$, \ $\epsilon\in(0,1)$, while $\abs\lambda^2$ is taken very high and  $v^2$ very small. In this limit $q_\infty$ \eqref{Epq} is big and one needs to shift the micromirror in such a way that the two optical paths in the MZI remain equal. Also the fluctuations of the position of the mirror \eqref{q2p2} become large. This could give a dispersion of the beam $C_1$, and also this effect should be corrected, say by the use of lenses.

By the choices \eqref{vsmall}, \eqref{strlaser},  from
\eqref{min}, \eqref{def:K}, \eqref{def:M} we get that $\Sigma_-(0)\big|_{\psi=\psi_0}$ is near its theoretical lower bound:
\begin{equation}\label{slb}
0<1+\Sigma_-(0)\big|_{\psi=\psi_0}\ll 1.
\end{equation}
By the connection \eqref{cmurepr} with the quadrature fluctuations, this means that a strong squeezing has been produced. To have also a good detection of this squeezing we need a strong cancellation of the shot noise in the spectrum of the difference current \eqref{S-final}; this means to have also  $\eta $ small, while the laser must bee sufficiently strong to keep  \eqref{strlaser} valid.

As one can check, by taking $\psi=\psi_1=\psi_0\pm \pi/2$ one obtains a very high maximum of $\Sigma_-(0)$. In \eqref{S-final}, the factor $\frac{c^2\varkappa^2\abs\lambda^2}{\mu^2 +\varkappa^2}$ appears; in the considered approximations $\abs\lambda^2$ is very strong, so, the constant $c^2$, coming from the detector response function, has to be taken sufficiently small in order to keep this factor finite.

It is worth noting that a good squeezing can be obtained also under less extreme conditions, with respect to \eqref{strlaser}. As an example, let us take
\begin{equation}\label{=2}
N(\om)=0, \qquad \frac\mgam\om=\frac 15, \qquad \frac{\eta\abs\lambda^2v^2}\Om=1.
\end{equation}
Then, \eqref{min} gives
\begin{equation}\label{S0-min}
\Sigma_-(0)\big|_{\psi_0}\simeq -0.773214 ,
\end{equation}
which indicates a strong squeezing.
By taking also $\eta=1/10$, we get also a good reduction of the shot noise in the detected beam
\[
1+(1-\eta)\Sigma_-(0)\big|_{\psi_0}\simeq 0.304108.
\]
By considering the complementary quadrature, i.e.\ $\psi_1=\psi_0\pm \pi/2$, we get instead
\begin{equation}\label{S0-max}
\Sigma_-(0)\big|_{\psi_1}\simeq 5.173214 .
\end{equation}

With the choice \eqref{=2} for the parameters, the conditions \eqref{vsmall} reduce to $v^2\ll 0.32$ and \eqref{=2} means to have a very strong laser. We expect the constant $v^2$ to be small and, perhaps, the laser has to be unrealistically strong. Moreover, by \eqref{q2p2}, the standard deviation of the position turns out to be of order 1; so, also in non extreme cases, the dispersion has to be corrected.

\begin{figure}[h]
\includegraphics[width=8.5truecm]{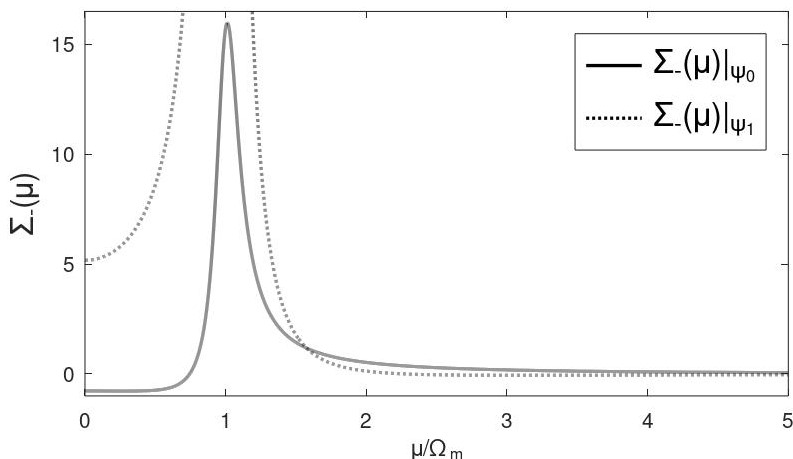}
\caption{The reduced spectrum $\Sigma_-(\mu)$ of the light reflected by the oscillating micromirror of Figure \ref{MZI}. The spectrum is plotted for the two different phases:  $\psi_0$, which minimizes $\Sigma_-(0)$ (see \eqref{min}) and $\psi_1=\psi_0\pm\pi/2$, which maximizes $\Sigma_-(0)$.  The frequency scale is in units of $\Om$, the bare frequency of the oscillator; the other parameters are taken as in \eqref{=2}. A negative value denotes squeezing of the corresponding quadrature \eqref{Qmu}.}\label{Sigma01}
\end{figure}

It is interesting to see the full reduced spectrum either for  phase $\psi_0$, which  minimizes $\Sigma_-(0)$, either for $\psi_1=\psi_0\pm \pi/2$, which maximizes it; these two cases are plotted in Fig.\ \ref{Sigma01} for the parameters \eqref{=2}.

\begin{figure}[h]
\includegraphics[width=8.5truecm]{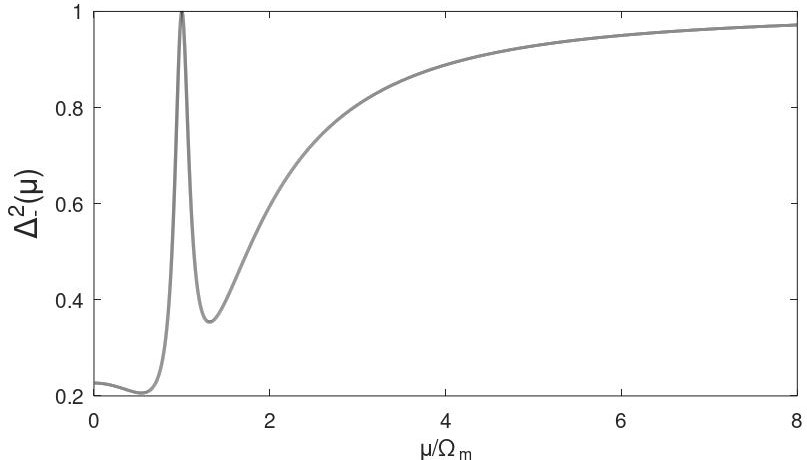}
\caption{The minimum variance $\Delta_-^2(\mu)$, \eqref{eqDelta-}, of the two-mode quadrature \eqref{Qmu} of the light reflected by the oscillating micromirror of Figure \ref{MZI}.  The other parameters are the same as in \eqref{=2}. A value under 1 denotes squeezing of at least a $\mu$-quadrature.}\label{Delta-}
\end{figure}

\begin{figure}[h]
\includegraphics[width=8.5truecm]{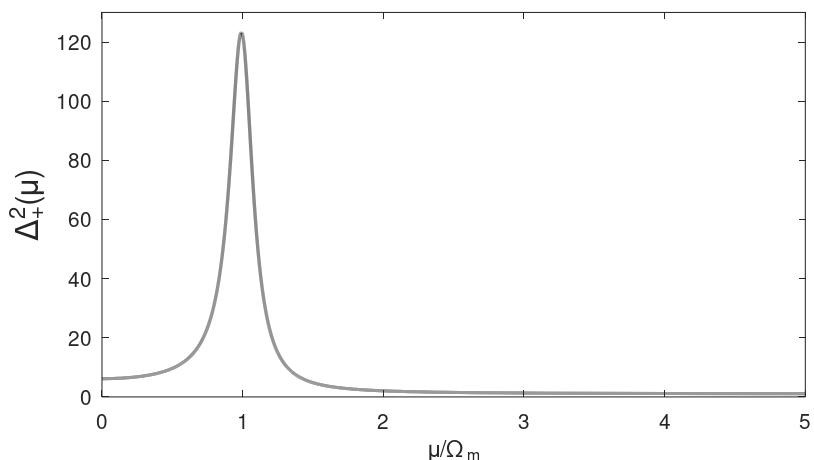}
\caption{The maximum variance $\Delta_+^2(\mu)$, \eqref{eqDelta+}, of the two-mode quadrature \eqref{Qmu} of the light reflected by the oscillating micromirror of Figure \ref{MZI}.  The other parameters are the same as in \eqref{=2}. With the choice of the phase maximizing the variance, the quadratures around $\mu/\Om=1$ appear to be strongly anti-squeezed.} \label{Delta+}
\end{figure}

We can see that $\Sigma_-(\mu)\big|_{\psi_0}$ is negative in a whole neighborhood of $\mu=0$ and, by enlarging the plot, that even $\Sigma_-(\mu)\big|_{\psi_1}$ becomes negative; so, squeezing does not concern a single mode. It is possible to quantify the squeezing by finding the minimum (and the maximum) for every $\mu$ of the quadrature fluctuations \eqref{Delta()}, which are connected to the spectrum by  \eqref{cmurepr}. So, we define
\begin{subequations}
\begin{gather}\label{eqDelta-}
\Delta^2_-(\mu)=\min_{\psi}\Delta^2(\mu,\psi),\\ \Delta^2_+(\mu)=\max_{\psi}\Delta^2(\mu,\psi).
\label{eqDelta+}
\end{gather}
\end{subequations}
Then, we can say that the variance of the two mode quadrature \eqref{Qmu}, for every fixed $\mu$, spans the interval $[\Delta^2_-(\mu),\Delta^2_+(\mu)]$ when the phase $\psi$ is varied.
The expression \eqref{Sigma-approx} can be easily minimized/maximized for every $\mu$; the result is
\begin{eqnarray}\nonumber
\Delta^2_\pm(\mu)&\simeq& 1+\frac{2\eta\abs\lambda^2v^2}{\Om\left[\left(1-\frac{\mu^2}\Omq\right)^2+4\left(1-\frac\omq\Omq\right)\frac{\mu^2}\Omq\right]} \\ {}&\times& \left\{E(\mu)\pm\sqrt{\left(1-\frac{\mu^2}\Omq\right)^2+E(\mu)^2}\right\}.
\end{eqnarray}
Note that $\Delta^2_-(\mu)<1$ for all the values of $\mu$, except for $\mu^2=1$, where it takes the value 1. The computations have been done under the conditions of Remark \ref{wint}; so, we see that there is squeezing for all choices of the parameters compatible with Remark \ref{wint} and for all $\mu$. For the choice of the parameters given in \eqref{=2}, the quantities $\Delta^2_\pm(\mu)$ are plotted in Figs.\ \ref{Delta-} and \ref{Delta+}; recall that the spectra are symmetric with respect to 0. An unexpected feature, shown by Fig.\ \ref{Delta-}, is that strong squeezing is present between  $\mu=0$ and $\mu=\pm 1$, and also around $\mu=\pm 1.5\Om$. Moreover, the high pick in Fig.\ \ref{Delta+} shows that around $\mu=\pm \Om$ the fluctuations are very far from the minimal uncertainty bound \eqref{UR}, i.e.\ $\Delta^2_-(\mu)\Delta^2_+(\mu)\geq 1$.

\subsection{The counting processes}
The  non-classical signature of the output light can be detected also by means of the counting processes at the two output ports and of the related Mandel $Q$-parameters \eqref{MQp}. Moreover,  a simple witness of squeezing is  the variance of the difference of counts; indeed, the parameter $Q_-$ is proportional to $\Sigma_-(0)$, see \eqref{MQp-}.

Let us consider the counting of photons at the two output ports of the MZI. The mean flux of counts are given in \eqref{n_jfinal}, and the Mandel parameters are obtained  from \eqref{MQpj}, \eqref{Sigma+0}, \eqref{def:M}, \eqref{def:theta}:
\begin{equation}\label{Mandelpars}
Q_j= \frac {\abs\lambda^2} { 4n_j}  \left[(-1)^j\sqrt{1-\eta}\,\Sigma_0(0)+\left(1-\eta\right)\Sigma_-(0)
\right].
\end{equation}
When $\Sigma_-(0) $ is negative, at least one of the Mandel parameters $Q_1$ and $Q_2$ is negative.
Moreover, when $\eta =0$ we get $Q_j=0$ from Eqs.\ \eqref{redsp}; indeed, the two counting processes are of Poisson type: the oscillator is not reached by the light and the input light is classical. When $\eta =1$ we get again  $Q_j=0$: the light interacts with the oscillator, but there is no interference with a reference beam and we only count the photons in the field  $C_1$ for which we have $\Lambda_{11}^C=\Lambda_{11}^B$, see \eqref{C=out}.

From \eqref{CovNN1}, \eqref{S12:cov3}, \eqref{Var+/T}--\eqref{MQp-} we have also
\begin{eqnarray*}
\lim_{T\to +\infty}&&\frac{\Cov_P[N_1(T),N_2(T)]}T=-\frac {\abs\lambda^2} { 4}  \left(1-\eta\right)\Sigma_-(0).
\\
\lim_{T\to+\infty}&&\frac{\Var_P[N_1(T)+ N_2(T)]}T=n_1+n_2=\abs\lambda^2,
\\
\lim_{T\to+\infty}&&\frac{\Var_P[N_1(T)- N_2(T)]}T=\abs\lambda^2\left[1+\left(1-\eta\right)\Sigma_-(0)\right],
\\
&&Q_+=0,\qquad Q_-=\left(1-\eta\right)\Sigma_-(0).
\end{eqnarray*}

Note that, for the sum of the counts we get the mean value  $\Ebb_P[N_1(T)+ N_2(T)]=\abs\lambda^2 T$ and the Mandel parameter $Q_+=0$.
Indeed, we have pure scattering on the oscillator, which gives rise to a phase change, without changing the total number of photons.

Under the extreme assumptions \eqref{vsmall}, \eqref{strlaser}, and for $\eta$ very small and the phase $\alpha_0$ introduced above, we find that the two Mandel parameters take a value slightly greater than $-1/2$. Similarly, under the same conditions, we find that the shot noise for the spectra \eqref{SIjfinal}  can be reduced nearly to half its value. So, as expected, the homodyne-like configuration and the measurement of the spectrum remains the most efficient way to detect squeezing in the field $C_1$.

\section{Conclusions and final remarks}\label{sec:concl}
In this work we have shown that the ideal apparatus of Figure \ref{MZI} could produce squeezed light out from a coherent input state (Sect.\ \ref{sqdetect}). This set up does not include any cavity; only travelling waves and pure scattering are involved. The squeezing is detected by counting measurements or by observing intensity spectra; in any case it appears as squeezing of some quadrature of the frequency mode operators \eqref{cmu}. When these operators involve a long time interval, they are strongly affected by the intensity dependent phase shifts of \eqref{Beminout}; instead, no squeezing can be seen in a short interval of time, because for nearly equal times the phase shifts compensate, as in \eqref{2Ls}.

In the case of very strong laser of Remark \ref{stronglaser}, the squeezing can be nearly total, as in \eqref{slb}; however, the assumption of extremely strong laser can be relaxed and a good squeezing can be reached under the condition \eqref{=2}. As we expect the parameter $v^2$ to be small, even this last condition could be problematic; in any case we have shown that, in principle, an oscillating quantum mirror can squeeze light by pure scattering, without the presence of a cavity.

Beyond the production of squeezed light, a second aim in our work was to show the flexibility of QSC and HP-equation; in constructing our example we have touched many points on open system theory, quantum optics, quantum information.

By the interaction of the phonon field with the quantum oscillator and the choice of field state (Sects.\ \ref{Sect:motion}--\ref{sec:equil}) we have shown how to describe a quantum damped \emph{mechanical} oscillator (with the classical dynamics, at least for mean values, see \eqref{eq:qp}) and how to model the interaction with a bath with arbitrary noise spectrum (non Markovian effects can be included, see also \cite{BarV15,Bar16}). Moreover, by using the scattering component of the HP-equation we have shown how to model the light scattering on the moving mirror, again generating the appearance of the right expression of the radiation pressure force in the Langevin equation for the momentum.

The HP-equation generates a unitary dynamics and the quantum Langevin equations are just the Heisenberg equations of motion for system operators. Having Bose fields, we can say that this construction realizes a unitary model of an open quantum system interacting with bosonic environments. Let us stress that what we are using can also be seen as a continuous-time limit of other approaches to Markovian and non-Markovian unitary dinamics, known as collision models or repeated interactions \cite{Meyer93,AttP06,Greg15,GiovSciarrMata19,Giov18,MEsp17,Brun02,Cicc17}.

Moreover, the formalism of QSC allows also to describe the monitoring of the system in continuous time. By introducing the fields in the Heisenberg description (the output fields, Sect.\ \ref{sec:IO}) one can individuate  combinations of these fields made  up of self-adjoint operators commuting at different times, and, so, representing compatible observables (Sects.\ \ref{sec:counts}, \ref{sec:pp&det}). By these means, also measurement based feedback can be introduced, and connections with quantum filtering, quantum trajectories, quantum stochastic master equations have been developed \cite{GJN12,BarG12,BarG13,WisM10,Car08,GarZ00,Bel12}.

Finally, we have shown how QSC and the generalized Weyl operators allow to describe a simple optical circuit, but more general linear optical devices can be modelled, such as polarizing beam splitters,  electro-optical phase modulators, wave platters\ldots
\cite{ASth,Bar06,BarG08b,BarG13,WisM10,BarG12,GJ09}. The choice of a Mach-Zehnder interferometer made in this work is due to the fact that it is a phase sensible apparatus; indeed, in a quantum context, it is often used in problems of estimation of a phase \cite{WisM10}. Moreover, as stressed in Sect.\ \ref{Sect:BS2}, when the whole apparatus is used to measure the spectrum of the difference current, it is similar to a homodyne scheme which is the usual way to detect the spectrum of squeezing \cite{Car08}.

\appendix

\section{Linear optical devices and Weyl operators}\label{lopt+W}

The generalized  Weyl operators are unitary operators on the Fock space $\Gamma$ \eqref{Focksp}, defined through their action on the exponential vectors \eqref{expV}:
\begin{eqnarray}\nonumber
&&\mathcal{W}(g;V)\in\Ucal(\Gamma), \quad  g\in L^2(\Rbb;\Cbb^d), \quad V\in \Ucal\big(L^2(\Rbb;\Cbb^d)\big),
\\ \label{Weyl}
&&\mathcal{W}(g;V)\, e(f) = \exp\left\{\rmi\IM \langle Vf|g\rangle \right\} e(Vf+g),
\end{eqnarray}
$\forall f\in L^2(\Rbb;\Cbb^d)$. From this definition  the following composition rule follows:
\begin{equation}\label{Weylprod}
\mathcal{W}(h;U)\,\mathcal{W}(g;V) =  \rme^{-\rmi \IM \langle h|Ug\rangle}
\mathcal{W}(h+Ug;UV).
\end{equation}

In the case $V=\openone$, it is possible to show that
\[
\Wcal(g;\openone)=\exp\biggl\{\sum_{k=1}^d\biggl(\int_{-\infty}^{+\infty}g_k(t)\rmd A_k^\dagger(t)- \text{h.c.}\biggr)\biggr\}.
\]
From \eqref{Weyl} one sees that $\Wcal(g;\openone)$ is the field analog of what is called a displacement operator in quantum optics \cite{BarG13}.

We shall need  Weyl operators
to analyze the output field dynamics in Sects.\ \ref{sec:IO}, \ref{sec:momWem}, \ref{app:th}, and also to describe linear optical elements in the MZI of Fig.\ \ref{MZI}. Indeed, when the Bose fields of QSC are used to describe travelling light waves, the Weyl operators can be used to describe linear optical elements \cite{ASth}.

Let $V$ be the unitary operator defined by $(Vf)_j(t)=\sum_i V_{ji}f_i(t)$, where $\{V_{ij}\}$ is a unitary matrix, $\sum_j \overline{V_{jk}}\,V_{ji}=\delta_{ki}$. Then, the Weyl operator $\mathcal{W}(0;V)$ gives the field transfomation
\begin{eqnarray}\nonumber
A_j(t) &\longmapsto & \mathcal{W}(0;V)^\dagger A_j(t) \mathcal{W}(0;V)=\sum_iV_{ji} A_i(t),
\\ \label{field+Weyl}
\Lambda_{ij}^A(t) &\longmapsto & \mathcal{W}(0;V)^\dagger \Lambda^A_{ij}(t) \mathcal{W}(0;V)
\\ &&{}\quad=\sum_{kl}\overline{V_{il}}\,V_{jk} \Lambda^A_{lk}(t).\nonumber
\end{eqnarray}
This transformation can be used to describe linear elements in optical circuits \cite{ASth}, such as (polarizing) beam splitters, half(or quarter)-wave platters, \ldots The key point is that these transformations, being unitary, preserve CCRs.

\paragraph{Beam splitter.} A
beam splitter of transmittance $\eta\in [0,1]$ can be represented by the Weyl operator $\mathcal{W}(0;V_\eta)$ with
\begin{equation}\label{Vbs}
V_{\eta}=\begin{pmatrix}\sqrt {\eta}& \rmi \sqrt{1-\eta }\\ \rmi \sqrt{1-\eta }& \sqrt {\eta}\end{pmatrix}.
\end{equation}
Inside the matrix $V_\eta$ different choices of the phases can be done. The  choice above is the one of \cite[Sect.\ 3.4]{ASth} and \cite[Sect.\ 14.4]{WalM94}. Another typical choice is to take $(V_\eta)_{12}=-\sqrt{1-\eta}=-(V_\eta)_{21}$ \cite[Eq.\ (5.9)]{Leo10}.  The different conventions are irrelevant because the physical phases can be adjusted by adding suitable phase shifts at the end.

\section{Computations of variances and spectra}\label{app:Var+sp}

For the detected fields we employ the usual notation of theoretical physics already introduced in Sect.\ \ref{Sect:oc}:
\begin{equation}\label{notation:d(t)}
d_j(t)=\frac{\rmd D_j(t)}{\rmd t},  \qquad \frac{\rmd \Lambda^D_{jj}(t)}{\rmd t} =d_j^\dagger(t)d_j(t);
\end{equation}
these fields satisfy the canonical commutation relations (CCRs):
\begin{equation}\label{d:ccr}
[d_j(t),d_i(s)]=0 ,\qquad [d_j(t),d_i^\dagger(s)]=\delta_{ij}\delta(t-s).
\end{equation}
In particular we have the key relation
\begin{eqnarray}\nonumber
d_j^\dagger(s)d_j(s)&&d_i^\dagger(t)d_i(t)=\delta_{ij}\delta(t-s)+d_j^\dagger(s)d_i^\dagger(t)d_j(s)d_i(t)\\ &&{}=\delta_{ij}\delta(t-s) +d_i^\dagger(t)d_j^\dagger(s)d_j(s)d_i(t).
\label{jjii:jiij}\end{eqnarray}
By using this relation inside the expression of the covariance \eqref{CovNN}, we obtain immediately \eqref{CovNN1}.

\subsection{The $d_j$-field expressions of the spectra}\label{dj:S}
Firstly, we have
\begin{eqnarray*}
\int_0^T\frac {\rme^{\rmi\mu t}}{\sqrt T}\,\hat I_j(t)\rmd t=\frac{c\varkappa}{\sqrt T}\int_0^T \rmd \Lambda_{jj}^D(r)\,\rme^{\varkappa r} \int_r^T \rmd t \,\rme^{\left(\rmi\mu -\varkappa\right)t}&&
\\ {}
{} =\frac{c\varkappa}{\left(\rmi \mu -\varkappa\right)\sqrt T}\int_0^T \rmd \Lambda_{jj}^D(t)\left[\rme^{\rmi \mu T -\varkappa\left(T-t\right)} -\rme^{\rmi\mu t}\right].&&
\end{eqnarray*}
Then, we introduce the field densities and we apply \eqref{jjii:jiij}:
\begin{eqnarray*}
\frac 1{T}&&\int_0^T\rmd t\int_0^T\rmd s\,\rme^{\rmi\mu (t-s)}\hat I_i(t)\hat I_j(s)
=\frac{c^2\varkappa^2}{\left(\mu^2 +\varkappa^2\right) T}
\\ &&{}\times \biggl\{\delta_{ij}
\int_0^T\rmd t \abs{\rme^{-\varkappa(T-t)}-\rme^{-\rmi\mu(T-t)}}^2d_j^\dagger(t)d_j(t)
\\ &&{}+\int_0^T \rmd t \int_0^T \rmd s \left[\rme^{-\varkappa\left(T-t\right)} -\rme^{-\rmi\mu \left(T-t\right)}\right]
\\ &&{}\times \left[\rme^{-\varkappa\left(T-s\right)} -\rme^{\rmi\mu \left(T-s\right)}\right]d_j^\dagger(t)d_i^\dagger(s) d_i(s)d_j(t)\biggr\}.
\end{eqnarray*}
Now we take the quantum expectation of the expression above, we subtract and add the term $\langle d^\dagger_j(t)d_j(t)\rangle_T $ $\times \langle d^\dagger_i(s)d_i(s)\rangle_T$ inside the second integral, and use $\langle d^\dagger_j(t)d_j(t)\rangle_T\simeq n_j$, which holds for $t$ large. In this way we get
\begin{eqnarray*}
\frac 1{T}&&\int_0^T\rmd t\int_0^T\rmd s\,\rme^{\rmi\mu (t-s)}\langle\hat I_i(t)\hat I_j(s)\rangle_T
\\  &&{}\simeq \frac{c^2\varkappa^2}{\left(\mu^2 +\varkappa^2\right) T}\biggl\{\delta_{ij}n_j\int_0^T\rmd t \abs{\rme^{-\varkappa(T-t)}-\rme^{-\rmi\mu(T-t)}}^2
\\ &&{} +n_in_j\biggl|\int_0^T \rmd t   \left[\rme^{-\varkappa\left(T-t\right)} -\rme^{-\rmi\mu \left(T-t\right)}\right]\biggr|^2
\\ &&{}+\RE\int_0^T \rmd t \int_0^T \rmd s \left[\rme^{-\varkappa\left(T-t\right)} -\rme^{-\rmi\mu \left(T-t\right)}\right]
\\ &&{}\times\left[\rme^{-\varkappa\left(T-s\right)} -\rme^{\rmi\mu \left(T-s\right)}\right]
\Bigl(\langle d_j^\dagger(t)d_i^\dagger(s) d_i(s)d_j(t)\rangle_T
\\ &&\quad {}-\langle d^\dagger_j(t)d_j(t)\rangle_T \langle d^\dagger_i(s)d_i(s)\rangle_T\Bigr)\biggr\}.
\end{eqnarray*}
By taking the limit for $T\to +\infty$ we obtain the equations \eqref{SI*2}, \eqref{Sigmaji}.

\subsection{The dependence on the output $c_1$-field}\label{sec:c1}
To compute the covariance matrix \eqref{CovNN1} and the spectra \eqref{Sigmaji}, we have to elaborate the quantity
\[
\langle d_j^\dagger(t)d_i^\dagger(s)d_i(s)d_j(t)\rangle_T
- \langle d_i^\dagger(s)d_i(s)\rangle_T\langle d_j^\dagger(t)d_j(t)\rangle_T.
\]
Let us take $t<T$, $ s<T$; from \eqref{d:Psi} we get
\begin{eqnarray*}
&&d_i(s)d_j(t)\, \rho_{\rm em}^{T}=\frac {\rmi^{i+j-2}} { 2}\Bigl[  c_1(s)c_1(t)
\\ {}&&+(-1)^j\rme^{\rmi \psi} \sqrt{1-\eta  }\,f(t)c_1(s) + (-1)^i\rme^{\rmi \psi} \sqrt{1-\eta  }\,f(s)c_1(t)
\\ {}&&\qquad \qquad {}+(-1)^{i+j}\rme^{2\rmi \psi}\left(1-\eta\right)f(s)f(t)\Bigr] \rho_{\rm em}^{T} .
\end{eqnarray*}
By this we obtain
\begin{eqnarray*}
&&\langle d_j^\dagger (t)d_i^\dagger (s)d_i(s)d_j(t)\rangle_T  - \langle d_j^\dagger (t)d_j(t)\rangle_T \langle d_i^\dagger (s)d_i(s)\rangle_T
\\ {}&&=\frac {1} { 4}\Bigl\{\langle c_1^\dagger(t)  c_1^\dagger(s)c_1(s)c_1(t)\rangle_T  -\langle c_1^\dagger(t)c_1(t)\rangle_T
\langle  c_1^\dagger(s)c_1(s)\rangle_T
\\ &&\quad {} +\Bigl[(-1)^j\rme^{\rmi \psi} \sqrt{1-\eta  }f(t)\Bigl(\langle c_1^\dagger(t)  c_1^\dagger(s)c_1(s) \rangle_T
\\ &&\quad {}-\langle c_1^\dagger(t)\rangle_T   \langle c_1^\dagger(s)c_1(s) \rangle_T \Bigr)
+ (-1)^i\rme^{\rmi \psi} \sqrt{1-\eta  }f(s)
\\ &&\quad {}\times\left(\langle c_1^\dagger(t)  c_1^\dagger(s)c_1(t)\rangle_T -\langle c_1^\dagger(s)\rangle_T \langle c_1^\dagger(t) c_1(t)\rangle_T \right)
\\ &&\quad {}+(-1)^{i+j}\rme^{2\rmi \psi}\left(1-\eta\right)f(s)f(t) \Bigl(\langle c_1^\dagger(t)  c_1^\dagger(s)\rangle_T
\\ &&\quad {}-\langle c_1^\dagger(s)\rangle_T \langle c_1^\dagger(t)  \rangle_T \Bigr)
+ (-1)^{i+j}\left(1-\eta\right)f(s)\overline{f(t)}
\\ &&\quad {}\times \left(\langle c_1^\dagger(s)c_1(t)  \rangle_T -\langle c_1^\dagger(s)\rangle_T  \langle c_1(t)  \rangle_T \right)
+ \text{c.c.}\Bigr]
\Bigr\}.
\end{eqnarray*}
In this expression the following quantities appear:
\begin{subequations}\label{Deltas}
\begin{eqnarray}\nonumber
\Delta_+(t,s;T)&:=&\langle c_1^\dagger(t)  c_1^\dagger(s)c_1(s)c_1(t)\rangle_T
\\ {}&-&\langle c_1^\dagger(t)c_1(t)\rangle_T \langle  c_1^\dagger(s)c_1(s)\rangle_T,
\label{Delta1}\end{eqnarray}
\begin{eqnarray}\nonumber
\Delta_0(t,s;T)&:=&f(t)\rme^{\rmi \psi} \Bigl(\langle c_1^\dagger(t)  c_1^\dagger(s)c_1(s) \rangle_T
\\ {}&-&\langle c_1^\dagger(t)\rangle_T   \langle c_1^\dagger(s)c_1(s) \rangle_T  \Bigr)
+ \text{c.c.},
\label{Delta2}\end{eqnarray}
\begin{eqnarray}\nonumber
&&\Delta_-(t,s;T):=\rme^{2\rmi \psi}f(s)f(t)\Bigl(\langle c_1^\dagger(t)  c_1^\dagger(s)\rangle_T
\\ \nonumber &&{}-\langle c_1^\dagger(s)\rangle_T \langle c_1^\dagger(t)  \rangle_T \Bigr)
+ f(s)\overline{f(t)}
\\ &&{} \times\left(\langle c_1^\dagger(s)c_1(t)  \rangle_T -\langle c_1^\dagger(s)\rangle_T  \langle c_1(t)  \rangle_T\right)+ \text{c.c.}.
\label{Delta3}\end{eqnarray}
\end{subequations}
Note that $\Delta_+(t,s;T)$ and $\Delta_-(t,s;T)$ are invariant for the exchange of $t$ and $s$.
By using these quantites we get
\begin{eqnarray}\nonumber
\langle d_j^\dagger &&(t)d_i^\dagger (s)d_i(s)d_j(t)\rangle_T  - \langle d_j^\dagger (t)d_j(t)\rangle_T \langle d_i^\dagger (s)d_i(s)\rangle_T
\\ \nonumber &&{}=\frac {1} { 4}\Bigl[\Delta_+(t,s;T)+(-1)^j\Delta_0(t,s;T)
\\ &&\ {}+(-1)^i\Delta_0(s,t;T)+(-1)^{i+j}\Delta_-(t,s;T)\Bigr].
\label{4j}\end{eqnarray}
Now we introduce the three quantities
\begin{subequations}\label{DeltatoSigma}
\begin{equation}
\Sigma_{\pm}(\mu):=\lim_{T\to+\infty}\frac 1 {\abs\lambda^2 T}\int_0^T \rmd t \int_0^T \rmd s \, \rme^{\rmi\mu \left(t-s\right)}\Delta_{\pm}(t,s;T),
\end{equation}
\begin{eqnarray}\nonumber
\Sigma_{0}(\mu):=\lim_{T\to+\infty}& &\frac 1 {\abs\lambda^2 T}\int_0^T \rmd t \int_0^T \rmd s \, \rme^{\rmi\mu \left(t-s\right)}\\ {}& &\times\left[\Delta_{0}(t,s;T)+\Delta_{0}(s,t;T)\right].
\end{eqnarray}
\end{subequations}
By the symmetry of the Delta-arguments and the fact that they are real, these quantities are real; then, also symmetric for $\mu \leftrightarrow -\mu$. The explicit forms in terms of $c_1$ are recalled in the main text, see \eqref{Sigmas}.

By inserting \eqref{4j} into $\Sigma_{ji}(\mu)$ \eqref{Sigmaji} we get \eqref{Sigma:c1}.

\section{Position and momentum fluctuations}\label{qpfl}
By direct computations of $\langle q(T)^2\rangle_T$ and $\langle p(T)^2\rangle_T$ from \eqref{q(t)}, \eqref{p(t)} and the properties of the field state, we get, in the limit $T\to +\infty$,
\begin{eqnarray}\nonumber
&&\langle q^2\rangle_{\rm eq}-q_\infty^2=\frac {\eta\abs\lambda^2v^2}{2\mgam}+ \frac{\Omq}{4\omq}\left(\frac{\abs x^2+\abs y^2}{\mgam} +\RE \frac {\rmi {\overline\tau}^3 xy}{2\Om}\right)
\\ &&{}+\int_\Rbb \rmd \nu \,\frac{\Omq N(\nu)}{4\omq\pi} \Biggl(\frac{\abs x^2}{\frac \mgamq 4 +\left(\om+\nu\right)^2} +\frac{\abs y^2}{\frac \mgamq 4 +\left(\om-\nu\right)^2}\nonumber
\\ &&{}-2\RE \frac{{\overline \tau}^2 xy}{ \left[\frac \mgam 2 +\rmi\left(\om +\nu\right)\right]\left[\frac \mgam 2 +\rmi\left(\om -\nu\right)\right]} \biggr), \label{momq2}
\end{eqnarray}
\begin{eqnarray}\nonumber
&&\langle p^2\rangle_{\rm eq}=\frac {\eta\abs\lambda^2v^2}{2\mgam}+ \frac{\Omq}{4\omq}\left(\frac{\abs x^2+\abs y^2}{\mgam} -\RE \frac {\rmi \overline\tau xy}{2\Om}\right)
\\ \nonumber
&&{}+\int_\Rbb \rmd \nu\,\frac{\Omq N(\nu)}{4\omq\pi}  \Biggl(\frac{\abs x^2}{\frac \mgamq 4 +\left(\om+\nu\right)^2} +\frac{\abs y^2}{\frac \mgamq 4 +\left(\om-\nu\right)^2}
\\ &&{}+2\RE \frac{ xy}{ \left[\frac \mgam 2 +\rmi\left(\om +\nu\right)\right]\left[\frac \mgam 2 +\rmi\left(\om -\nu\right)\right]} \biggr),
\label{momp2}
\end{eqnarray}
\[
x:=\tau\alpha +\rmi \beta,\qquad y:= \tau \alpha +\rmi \overline \beta.
\]

To get the equality of \eqref{momq2} with \eqref{momp2} when  $N(\nu)$ is arbitrary, we need the equality of the integrands. As ${\overline \tau}^2 \neq -1$, we must have $xy=0$. By the conditions \eqref{constants}, we have $y\neq 0$. Thus, we obtain  $x=0$, which implies \eqref{Rfinal} and  $y=\sqrt{\frac{2\mgam \om}\Om}$. By inserting these results into \eqref{momq2}, \eqref{momp2} we get \eqref{q2p2}, with $N_{\rm eff}$ defined in \eqref{Neff}; then, again by direct computations, we get \eqref{q,p}.

\section{Proof of Proposition \ref{proposition}}\label{proof:prop}
By using the decomposition \eqref{SWW} and the composition rules of the Weyl operators \eqref{Weylprod} we get
\begin{subequations}\label{2decomps}
\begin{eqnarray}\nonumber
\rme^{\rmi v q(t)}\rme^{\pm\rmi v q(s)}&=&\rme^{\pm \rmi \IM\langle \ell_s|\ell_t\rangle}S_0(t)S_0(s)^\pm
\\ {}&\times& W_1(V_s^\pm V_t)W_3(\ell_t\pm \ell_s),
\\ \nonumber
\rme^{\pm\rmi v q(s)}\rme^{\rmi v q(t)}&=&\rme^{\mp \rmi \IM\langle \ell_s|\ell_t\rangle}S_0(s)^\pm S_0(t)
\\ {}&\times& W_1(V_s^\pm V_t)W_3(\ell_t\pm \ell_s).
\end{eqnarray}
\end{subequations}

We have $c_1(s)c_1(t)=c_1(t)c_1(s)$, \ $V_tV_s=V_sV_t$. \ By using \eqref{SWW} and \eqref{Weyl} we obtain
\begin{eqnarray*}
&&c_1(t)c_1(s)e^B_1(f_1)\otimes e_3^A(f_3)=\rme^{\rmi v q(t)+\rmi \phi}b_1(t)\rme^{\rmi v q(s)+\rmi \phi}
b_1(s)
\\ &&{}\times e^B_1(f_1)\otimes e_3^A(f_3)
=\rme^{\rmi v q(t)+2\rmi \phi}b_1(t)S_0(s)\rme^{-\rmi\IM \langle \ell_s|f_3\rangle}
\\ &&{}\times f_1(s)e^B_1(V_sf_1)\otimes e_3^A(f_3+\ell_s)
=
\rme^{2\rmi\phi}f_1(s)\rme^{-\rmi\IM \langle \ell_s|f_3\rangle}
\\ &&{}\times S_0(t)S_0(s)V_s(t)f_1(t)\rme^{-\rmi\IM \langle \ell_t|f_3+\ell_s\rangle}e^B_1(V_tV_sf_1)
\\ &&{}\otimes e_3^A(f_3+\ell_s+\ell_t)
=
\rme^{-\rmi\IM \langle \ell_t|\ell_s\rangle}V_s(t)S_0(t)S_0(s)f_1(t)
\\ &&{}\times f_1(s)\rme^{2\rmi\phi-\rmi\IM \langle\ell_t+ \ell_s|f_3\rangle}e^B_1(V_tV_sf_1)\otimes e_3^A(f_3+\ell_s+\ell_t),
\end{eqnarray*}
\begin{eqnarray*}
c_1(s)&&c_1(t)e^B_1(f_1)\otimes e_3^A(f_3)=
\rme^{-\rmi\IM \langle \ell_s|\ell_t\rangle}V_t(s)
\\ &&{}\times S_0(s)S_0(t)f_1(t)f_1(s)\rme^{2\rmi\phi-\rmi\IM \langle\ell_t+ \ell_s|f_3\rangle}
\\ &&{}\times e^B_1(V_tV_sf_1)\otimes e_3^A(f_3+\ell_s+\ell_t).
\end{eqnarray*}
By the arbitrariness of the coherent state one has \eqref{commutS_0}. By using \eqref{2decomps} and \eqref{commutS_0} we get \eqref{Vqq}. Eq.\ \eqref{Vqqq} follows trivially from \eqref{Vqq}. Equality \eqref{[,]} is proved by using the factorization and applying the operators in the two sides to a generic coherent vector.

\section{Detected fields and scattering operator}\label{Dmom_comp}

By inserting the expression \eqref{keyeq} into \eqref{meanN3} and \eqref{Deltas}, we get immediately
\begin{equation}\label{nj:q}
n_j=\lim_{t\to+\infty} \lim_{T\to+\infty} \frac12 \left[\abs\lambda^2  + (-1)^j  \chi \RE\rme^{\rmi \left(\phi-\psi\right)}\langle\rme^{\rmi v q(t)}\rangle_T\right],
\end{equation}
\begin{equation}\label{Delta1c1}
\Delta_+(t,s;T)=0 \quad \Rightarrow \quad \Sigma_+(\mu)=0,
\end{equation}
\begin{eqnarray*}
\Delta_0(t,s;T)&=&\rme^{\rmi \left(\psi-\phi\right)} \eta \abs\lambda^2\Bigl( f(s)\langle \rme^{-\rmi vq(t)} b_1^\dagger(s) \rangle_T \\ {}&-&\abs\lambda^2\sqrt\eta \langle \rme^{-\rmi vq(t)} \rangle_T \Bigr) + \text{c.c.},
\end{eqnarray*}
\begin{eqnarray*}
&&\Delta_-(t,s;T)=\abs\lambda^2\rme^{2\rmi \left(\psi-\phi\right)}
\\ &&\ {}\times \Bigl(f(s)\sqrt\eta \langle \rme^{-\rmi vq(t)} b_1^\dagger(s)  \rme^{-\rmi vq(s)}\rangle_T
\\ &&\ {} -\eta\abs\lambda^2\langle \rme^{-\rmi vq(s)}\rangle_T \langle \rme^{-\rmi vq(t)} \rangle_T \Bigr)
+ \eta\abs\lambda^4
\\ \ &&{}\times\left(\langle \rme^{-\rmi vq(s)}\rme^{\rmi vq(t)} \rangle_T -\langle \rme^{-\rmi vq(s)}\rangle_T\,  \langle \rme^{\rmi vq(t)} \rangle_T\right)+ \text{c.c.}.
\end{eqnarray*}
Then, by using \eqref{[,]} and its adjoint, we get
\begin{equation*}
\Delta_0(t,s;T)=\rme^{\rmi \left(\phi-\psi\right)}  \abs\lambda^4\eta^{3/2}  \left(V_t(s)-1\right)\langle \rme^{\rmi vq(t)} \rangle_T   + \text{c.c.},
\end{equation*}
\begin{eqnarray*}
\Delta_-&&(t,s;T)=\eta\abs\lambda^4\Bigl[\rme^{2\rmi \left(\phi-\psi\right)}\Bigl(V_t(s)\langle  \rme^{\rmi vq(s)}\rme^{\rmi vq(t)}\rangle_T
\\ &&{}-\langle \rme^{\rmi vq(s)}\rangle_T \langle \rme^{\rmi vq(t)} \rangle_T \Bigr)
\\ &&{} + \langle \rme^{-\rmi vq(s)}\rme^{\rmi vq(t)} \rangle_T -\langle \rme^{-\rmi vq(s)}\rangle_T\,  \langle \rme^{\rmi vq(t)} \rangle_T\Bigr]+ \text{c.c.}.
\end{eqnarray*}
Finally, by using also \eqref{Vts}, the reduced spectra \eqref{DeltatoSigma} become
\begin{eqnarray}\nonumber
\Sigma_0&&(\mu) = \lim_{T\to+\infty}\frac {2\eta^{3/2}\abs\lambda^2} { T}\,\RE\int_0^T \rmd t \int_0^t \rmd s \, \rme^{\rmi\mu s}
\\ &&{}\times\Bigl\{\rme^{\rmi \left(\phi-\psi\right)}   \left(\rme^{2\rmi h(s)}-1\right)\langle \rme^{\rmi vq(t)} \rangle_T
+ \text{c.c.}\Bigr\},
\label{Sigma0q}\end{eqnarray}
\begin{eqnarray}\nonumber
&&\Sigma_-(\mu) = \lim_{T\to+\infty}\frac {\eta\abs\lambda^2} { T}\int_0^T \rmd t \int_0^T \rmd s \, \rme^{\rmi\mu \left(t-s\right)}
\Bigl\{ \rme^{2\rmi \left(\phi-\psi\right)}
\\ \nonumber
{}&&\times \Bigl[V_t(s)\langle  \rme^{\rmi vq(s)}\rme^{\rmi vq(t)}\rangle_T
-\langle \rme^{\rmi vq(s)}\rangle_T \langle \rme^{\rmi vq(t)} \rangle_T \Bigr]
\\ \nonumber
{}&&+ \langle \rme^{-\rmi vq(s)}\rme^{\rmi vq(t)} \rangle_T -\langle \rme^{-\rmi vq(s)}\rangle_T\,  \langle \rme^{\rmi vq(t)} \rangle_T+ \text{c.c.}\Bigr\}
\\ \nonumber
{}&&= \lim_{T\to+\infty}\frac {2\eta\abs\lambda^2} { T}\,\RE\int_0^T \rmd t \int_0^t \rmd s \, \rme^{\rmi\mu s} \Bigl\{ \rme^{2\rmi \left(\phi-\psi\right)}
\\ \nonumber &&
{}\times\Bigl[\rme^{2\rmi h(s)}\langle  \rme^{\rmi vq(t-s)}\rme^{\rmi vq(t)}\rangle_T
-\langle \rme^{\rmi vq(t-s)}\rangle_T \langle \rme^{\rmi vq(t)} \rangle_T \Bigr]
\\ \nonumber && \quad
{}+ \langle \rme^{-\rmi vq(t-s)}\rme^{\rmi vq(t)} \rangle_T \\ &&\qquad {}-\langle \rme^{-\rmi vq(t-s)}\rangle_T\,  \langle \rme^{\rmi vq(t)} \rangle_T+ \text{c.c.}\Bigr\}.
\label{Sigma-q}\end{eqnarray}

\subsection{Moments of the electromagnetic Weyl operator}\label{sec:momWem}
Let us consider the Weyl operator $W_1(V_t)$ \eqref{Wem} with $V_t$ defined by \eqref{Vts}, \eqref{V(t)}; it involves only the electromagnetic field $B_1$.
Let us take now $0<s<t\leq T$ and let $T$ first and $t$ after go to $+\infty$;
by the definition \eqref{Weyl} and the composition rule for Weyl operators \eqref{Weylprod} we get
\begin{subequations}\label{<1>}
\begin{eqnarray}
&&\langle W_1(V_t)\rangle_T=\exp\left\{\eta \abs{\lambda}^2\int_0^t\rmd s  \left(\rme^{2\rmi h(s)} -1\right)\right\}
\\ &&{}\simeq \exp\left\{ \eta \abs{\lambda}^2\int_0^{+\infty}\rmd s  \left(\rme^{2\rmi h(s)} -1\right)\right\}
=\rme^{-M +\rmi \theta},\nonumber
\end{eqnarray}
\begin{equation}
\langle W_1(V_t^\dagger)\rangle_T\simeq \rme^{-M -\rmi \theta};
\end{equation}
\end{subequations}
the quantities $M$ and $\theta$ are defined in \eqref{def:M}, \eqref{def:theta}. Then, we have also
\begin{widetext}
\begin{eqnarray}\nonumber
\langle W_{1}(V_{t-s}V_t)\rangle_T&=&\exp\left\{\eta  \langle f_T|\left(V_{t-s} V_t -\openone\right)f_T\rangle\right\}
\\ \nonumber {}&=&
\exp\left\{\eta  \langle f_T|\left(V_{t-s}-\openone \right)\left(V_t-\openone \right)f_T\rangle+ \eta  \langle f_T|\left(V_{t-s}-\openone \right)f_T\rangle+ \eta  \langle f_T|\left(V_t-\openone \right)f_T\rangle\right\}
\\ \nonumber {}&=& \exp\left\{\eta \abs\lambda^2\int_0^{t- s}\rmd u \left(\rme^{2\rmi h(t-s-u)} -1\right) \left(\rme^{2\rmi h(t-u)} -1\right) \right\}\langle W_1(V_{t-s})\rangle_T\langle W_1(V_t)\rangle_T
\\ \nonumber {}&=& \exp\left\{\eta \abs\lambda^2\int_0^{t-s}\rmd u \left(\rme^{2\rmi h(u)} -1\right) \left(\rme^{2\rmi h(s+u)} -1\right) \right\}\langle W_1(V_{t-s})\rangle_T\langle W_1(V_t)\rangle_T
\\ {}&\simeq&  \exp\left\{\eta \abs\lambda^2\int_0^{+\infty}\rmd u \left(\rme^{2\rmi h(u)} -1\right) \left(\rme^{2\rmi h(s+u)} -1\right) \right\} \rme^{-2M+2\rmi\theta},\label{VV+}
\end{eqnarray}
\begin{eqnarray}\nonumber
\langle W_1(V_{t-s}^\dagger)W_1(V_t)\rangle_T&=&\exp\left\{\eta \abs\lambda^2\int_0^{t-s}\rmd u \left(\rme^{-2\rmi h(u)} -1\right) \left(\rme^{2\rmi h(s+u)} -1\right) \right\}\langle W_1(V_{t-s}^\dagger)\rangle_T\langle W_1(V_t)\rangle_T
\\ {}&\simeq & \exp\left\{\eta \abs\lambda^2\int_0^{+\infty}\rmd u \left(\rme^{-2\rmi h(u)} -1\right) \left(\rme^{2\rmi h(s+u)} -1\right) \right\} \rme^{-2M}.\label{VV}
\end{eqnarray}

\subsection{Moments of the thermal Weyl operator}\label{app:th}
Let us consider now the thermal Weyl operator $W_3(\ell_t)$ \eqref{Wth}, with $\ell_t$ defined in \eqref{ell(t)}; recall that the thermal state \eqref{uTst} is a mixture of the coherent states $e_{3}(u_T)$. Then, we have
\begin{eqnarray*}
\langle e_{3}(u_T)&&|W_3(\ell_t)|e_{3}(u_T)\rangle=\rme^{\rmi \IM \langle u_T|\ell_t\rangle}\langle e_3(u_T)|e_3(u_T+\ell_t)\rangle
\\ &&{}=\exp\biggl\{\rmi \IM \langle u_T|\ell_t\rangle-\frac 12 \norm{u_T}^2 -\frac 12 \norm{u_T+\ell_t}^2+\langle u_T|u_T+\ell_t\rangle\biggr\}
=\exp\left\{\langle u_T|\ell_t\rangle- \langle \ell_t|u_T\rangle  -\frac 12 \norm{\ell_t}^2\right\}.
\end{eqnarray*}
Being $u_T$ a Gaussian process with moments \eqref{momu}, we have
\begin{eqnarray*}\nonumber
\langle W_3(\ell_t) \rangle_T&=& \exp\left\{ -\frac 12 \norm{\ell_t}^2- \int_0^T\rmd s \int_0^T \rmd r \, \ell_t(s)F(s-r)\, \overline{\ell_t(r)}\right\}
\\ \nonumber {} &=& \exp\left\{ - \frac{\Om v^2}{4\om}\left(1-\rme^{-\mgam t}\right)- \frac 1{2\pi}\int_{-\infty}^{+\infty}\rmd \nu\, N(\nu) \abs{ \int_0^T \rmd s\, \ell_t(s)\rme^{\rmi \nu s}}^2\right\}
\\ \nonumber {}&=&\exp\biggr\{-\frac{\Om v^2}{4\om}\left(1-\rme^{-\mgam t}\right)-\frac{\mgam\Om v^2}{4\pi\om} \int_{-\infty}^{+\infty}\rmd \nu \,\frac{ N(\nu)}{ (\nu-\om)^2+\frac{\mgam^2}4}  \abs{\rme^{\rmi(\nu-\om)t}-\rme^{-\frac\mgam 2\,t}}^2\biggr\},
\end{eqnarray*}
\begin{equation}
\langle W_3(\ell_t) \rangle_T\simeq \exp\left\{-\frac{\Om v^2}{2\om}
\left(N_{\rm eff}+\frac 1 2\right)\right\}=\rme^{-K}, \label{3mI}
\end{equation}
where $K$ is defined by \eqref{def:K}, \eqref{Neff}.
In a similar way we get
\begin{eqnarray*}
\abs{ \int_0^T \rmd u \left( \ell_t(u)\pm\ell_{t-s}(u)\right)\rme^{\rmi \nu u}}^2&=& \frac{\Om \mgam  v^2}{2\om}
\abs{\int_0^t \rmd u \,\rme^{\left(\rmi \om -\frac \mgam  2\right)(t-u)+\rmi \nu u}\pm \int_0^{t-s} \rmd u \,\rme^{\left(\rmi \om -\frac \mgam  2\right)(t-s-u)+\rmi \nu u}}^2
\\ {}&\simeq& \frac{\Om \mgam  v^2}{2\om\left[\left(\nu-\om\right)^2+\frac{\mgamq}4\right]}\abs{\rme^{\rmi\nu t}\pm \rme^{\rmi\nu (t-s)}}^2= \frac{\Om \mgam  v^2}{\om\left[\left(\nu-\om\right)^2+\frac{\mgamq}4\right]}\left[1\pm \cos\nu s\right],
\end{eqnarray*}
\begin{eqnarray}\nonumber
\langle W_3(\ell_t&\pm&\ell_{t-s})\rangle_T=  \exp\biggl\{ -\frac 12 \norm{\ell_t\pm\ell_{t-s}}^2- \frac 1{2\pi}\int_{-\infty}^{+\infty}\rmd \nu\, N(\nu) \abs{ \int_0^T \rmd u \left( \ell_t(u)\pm\ell_{t-s}(u)\right)\rme^{\rmi \nu u}}^2\biggr\}
\\ {}&\simeq& \exp\biggl\{-\frac {v^2\Om}{2\om}\biggl[ 1 \pm \rme^{-\frac\mgam 2 \,s}\cos\om s + \int_\Rbb \rmd \nu\, \frac{\mgam N(\nu)\left[1\pm \cos\nu s\right]}{\pi\left[ \left(\nu-\om\right)^2+\mgam^2/4\right]} \biggr]\biggr\}\simeq \exp\left\{-2K \mp g(s) \right\};\label{W3W3}
\end{eqnarray}
$g(s)$ is defined in \eqref{def:g}.

\subsection{Moments of the scattering operator}\label{mainmoments}
For $s$ and/or $t$ large, we get $\IM\langle \ell_s|\ell_t\rangle \simeq h(t-s) $ and
\begin{equation}\label{prodscatt}
\rme^{\pm\rmi v q(s)}\rme^{\rmi v q(t)}\simeq \rme^{\mp \rmi h(t-s)}W_1(V_s^\pm V_t)W_3(\ell_t\pm \ell_s).
\end{equation}

In the following we take $0<s<t\leq T$, and we consider $T$ and $t$ large. Firstly we have
\begin{equation}\label{for_mean}
\langle\rme^{\rmi v q(t)}\rangle_T\simeq \rme^{-(K+M) +\rmi \theta}.
\end{equation}
By inserting this expression into \eqref{Sigma0q} we get the reduced spectrum \eqref{Sigma+0}.
By \eqref{prodscatt}, \eqref{VV+}, \eqref{VV}, \eqref{W3W3}, we obtain
\begin{eqnarray*}
\langle\rme^{\pm\rmi v q(t-s)}\rme^{\rmi v q(t)}\rangle_T&\simeq& \rme^{\mp \rmi h(s)}\langle W_1(V_{t-s}^\pm V_t)W_3(\ell_t\pm \ell_{t-s})\rangle_T
\\  \ {}&\simeq &\rme^{-2(K+M)\mp g(s)\mp \rmi h(s)} \, \rme^{\rmi \left(\theta\pm \theta\right)}
\exp\left\{\eta \abs\lambda^2\int_0^{+\infty}\rmd u \left(\rme^{\pm 2\rmi h(u)} -1\right) \left(\rme^{2\rmi h(s+u)} -1\right) \right\},
\end{eqnarray*}
\begin{eqnarray*}
\langle \rme^{- \rmi v q(t-s)}\rme^{\rmi v q(t)}\rangle_T &&- \langle \rme^{- \rmi v q(t-s)}\rangle_T\langle\rme^{\rmi v q(t)}\rangle_T
\\ &&{} \simeq  \rme^{-2(K+M)}
\biggl[\exp\Bigl\{ g(s) +\rmi h(s)
+\eta \abs\lambda^2\int_0^{+\infty}\rmd u \left(\rme^{- 2\rmi h(u)}-1\right)\left( \rme^{2\rmi h(s+u)} -1\right)\Bigr\}-1\biggr],
\end{eqnarray*}
\begin{eqnarray}\nonumber
\langle \rme^{- \rmi v q(t-s)}\rme^{\rmi v q(t)}\rangle_T &&- \langle \rme^{- \rmi v q(t-s)}\rangle_T\langle\rme^{\rmi v q(t)}\rangle_T
+\rme^{2\rmi \left(\phi-\psi\right)}
\left(\rme^{2\rmi h(s)}\langle \rme^{\rmi v q(t-s)}\rme^{\rmi v q(t)}\rangle_T-\langle \rme^{ \rmi v q(t-s)}\rangle_T\langle\rme^{\rmi v q(t)}\rangle_T\right)
\\ \nonumber &&{}\simeq \rme^{-2(K+M)}
\biggl(\biggl[\exp\Bigl\{ g(s) +\rmi h(s)
+\eta \abs\lambda^2\int_0^{+\infty}\rmd u \left(\rme^{- 2\rmi h(u)}-1\right)\left( \rme^{2\rmi h(s+u)} -1\right)\Bigr\}-1\biggr]
\\ &&{}+\rme^{-2\rmi\alpha}\biggl[\exp\Bigl\{
+\eta \abs\lambda^2\int_0^{+\infty}\rmd u \left(\rme^{2\rmi h(u)}-1\right)
\left( \rme^{2\rmi h(s+u)} -1\right)
- g(s) +\rmi h(s)\Bigr\}-1\biggr]\biggr),
\label{xdiff}\end{eqnarray}
with $\alpha$ defined in \eqref{def:K}. By inserting this expression into \eqref{Sigma-q} we get \eqref{compSigma0}, \eqref{compSigmapsi}.

\section{Computations of the approximated expressions for the reduced spectra}\label{sec:approx}

By \eqref{def:h(t)} and the assumptions \eqref{vsmall},  we obtain
\begin{eqnarray*}
\eta \abs\lambda^2&\RE&\int_0^{+\infty}\rmd u \left(\rme^{\mp 2\rmi h(u)}-1\right)\left( \rme^{ 2\rmi h(s+u)} -1\right)
\\
{}&\simeq &\pm\eta \abs\lambda^2\int_0^{+\infty}\rmd u \, 4 h(u) h(s+u)
=\pm\frac{\eta\abs\lambda^2 v^4\Om}{4\mgam\om }\Bigl[\tau\rme^{\left(\rmi \om -\frac\mgam 2\right)s}+\text{c.c.}\Bigr].
\end{eqnarray*}
Then, from \eqref{compSigma0}, we get
\begin{eqnarray*}
&&\Sigma_-^0(\mu)\simeq 2\abs\lambda^2\eta\rme^{-2(K+M)} \RE\int_0^{+\infty} \rmd s\, \rme^{\rmi\mu s}\biggl\{
g(s) +\rmi h(s)
+\frac{\eta\abs\lambda^2 v^4\Om}{4\mgam\om }\Bigl[\tau\rme^{\left(\rmi \om -\frac\mgam 2\right)s}+\text{c.c.}\Bigr]
+\text{c.c.}\biggr\}
\\ &&{}=
2\abs\lambda^2\eta\rme^{-2(K+M)} \frac{v^2\Om}{2\om}\RE\biggl\{
\frac{2N(\om)+1}{\frac\mgam 2 -\rmi \left(\mu+\om\right)}+\frac{2N(\om)+1}{\frac\mgam 2 -\rmi \left(\mu-\om\right)}
+\frac{\eta\abs\lambda^2 v^2}{\mgam }
\Bigl[\frac{\tau}{\frac\mgam 2 -\rmi \left(\mu+\om\right)}
\\ &&{}+\frac{\overline\tau}{\frac\mgam 2 -\rmi \left(\mu-\om\right)}\Bigr]\biggr\}=
2\abs\lambda^2\eta\rme^{-2(K+M)} \frac{v^2\Om}{2\om}
\biggl\{\frac\mgam 2 \biggl(
\frac{2N(\om)+1}{\frac\mgamq 4+ \left(\mu+\om\right)^2}
+\frac{2N(\om)+1}{\frac\mgamq 4 + \left(\mu-\om\right)^2}\biggr)
+\frac{\eta\abs\lambda^2 v^2}{2\Om }
\\ && {}\times\Bigl[\frac{2\om+\mu}{\frac\mgamq 4 + \left(\mu+\om\right)^2}+\frac{2\om-\mu}{\frac\mgamq 4 + \left(\mu-\om\right)^2}\Bigr]\biggr\}
= \eta\abs\lambda^2v^2
\Om\rme^{-2(K+M)}
\,\frac{\frac\mgam\om\left(2N(\om)+1\right)\left(\Omq+\mu^2\right)+2\eta\abs\lambda^2 v^2\Om}{\left[\frac\mgamq 4 + \left(\mu+\om\right)^2\right]\left[\frac\mgamq 4 + \left(\mu-\om\right)^2\right]}.
\end{eqnarray*}
Similarly, from \eqref{compSigmapsi}, we get
\begin{eqnarray*}
\Sigma_-^\psi(\mu)&\simeq& 2\abs\lambda^2\eta\rme^{-2(K+M)} \RE\int_0^{+\infty} \rmd s\, \rme^{\rmi\mu s}\Bigl\{\rme^{-2\rmi \alpha} \Bigl[-g(s)+\rmi h(s) - \frac{\eta\abs\lambda^2 v^4\Om}{4\mgam\om }
\Bigl(\tau\rme^{\left(\rmi \om -\frac\mgam 2\right)s}+\text{c.c.}\Bigr)\Bigr]+\text{c.c.}\Bigr\}
\\ {}&=&-\frac{\abs\lambda^2\eta v^2\Om}\om\,\rme^{-2(K+M)}
\RE\biggl\{\frac 1 {\frac\mgam 2-\rmi\left(\mu-\om\right)}
\biggl[N(\om)\rme^{2\rmi\alpha}
+\left(N(\om)+1\right)\rme^{-2\rmi\alpha}+\frac{\eta\abs\lambda^2v^2}\mgam\, \overline\tau\cos2\alpha\biggr]
\\ {}&+&\frac 1 {\frac\mgam 2-\rmi\left(\mu+\om\right)}\biggl[N(\om)\rme^{-2\rmi\alpha}
+\left(N(\om)+1\right)\rme^{2\rmi\alpha}+\frac{\eta\abs\lambda^2v^2}\mgam\, \tau\cos2\alpha\biggr]\biggr\}
=-\frac{\abs\lambda^2\eta v^2\Om}\om\,\rme^{-2(K+M)}
\\ {}&\times&
\biggl\{\frac 1 {\frac\mgamq 4+\left(\mu-\om\right)^2}
\biggl[\frac \mgam 2 \biggl(2N(\om)+1 +\frac{\eta\abs\lambda^2v^2\om}{\mgam\Om}\biggr) \cos2\alpha
+\left(\mu-\om\right)\biggl(\sin 2 \alpha- \frac{\eta\abs\lambda^2v^2}{2\Om}\cos2\alpha\biggr)\biggr]
\\ {}&+&\frac 1 {\frac\mgamq 4+\left(\mu+\om\right)^2}
\biggl[\frac \mgam 2 \biggl(2N(\om)+1 +\frac{\eta\abs\lambda^2v^2\om}{\mgam\Om}\biggr) \cos2\alpha
-\left(\mu+\om\right)\biggl(\sin 2 \alpha- \frac{\eta\abs\lambda^2v^2}{2\Om}\cos2\alpha\biggr)\biggr]\biggr\}
\\ &=& \eta\abs\lambda^2v^2
\Om\rme^{-2(K+M)}
\,\frac{2\left(\Omq -\mu^2\right) \sin 2\alpha -\left(2\eta\abs\lambda^2 v^2\Om+\frac\mgam \om\left(2N(\om)+1\right)\left(\Omq+\mu^2\right)\right)\cos2\alpha}{\left[\frac\mgamq 4 + \left(\mu+\om\right)^2\right]\left[\frac\mgamq 4 + \left(\mu-\om\right)^2\right]}.
\end{eqnarray*}
By summing up these two results and taking into account $\rme^{-(K+M)}\simeq 1$,  \eqref{Sigma-approx} follows.
Finally, from \eqref{Sigma+0}, we have
\begin{eqnarray*}
\Sigma_0(\mu)&\simeq& 2\rme^{-(K+M)}\eta^{3/2}\abs\lambda^2
\RE\int_0^{+\infty} \rmd t\, \rme^{\rmi\mu t}
\left[\rme^{-\rmi \alpha}2\rmi h(t)
+\text{c.c.} \right]
\\ {}&=&\rme^{-(K+M)}\,\frac{\eta^{3/2}\abs\lambda^2v^2\Om}\om
\RE\biggl[\biggl(\frac 1 {\frac\mgam 2 -\rmi \left(\mu-\om\right)}-\frac 1 {\frac\mgam 2 -\rmi \left(\mu+\om\right)}\biggr)2\rmi\sin\alpha
\biggr] ,
\end{eqnarray*}
from which we get \eqref{Sigma0approx}.
\end{widetext}


\begin{thebibliography}{99}
\bibitem{JacTWS99} K. Jacobs, I. Tittonen, H.M. Wiseman, and S. Schiller, \textsl{Quantum noise in the position measurement of a cavity mirror undergoing Brownian motion}, Phys. Rev. A \textbf{60} (1999) 538--548.

\bibitem{GioV01} V. Giovannetti and D. Vitali, \textsl{Phase noise measurement in a cavity with a movable mirror undergoing quantum Brownian motion}, Phys. Rev. A \textbf{63}  (2001) 023812.
\bibitem{GMVT09} C. Genes, A. Mari, D. Vitali, and P. Tombesi, \textsl{Quantum Effects in Optomechanical Systems}, Adv. At. Mol. Opt. Phys. \textbf{57} (2009) 33--86 .
\bibitem{Vitali12} M. Abdi, A.R. Bahrampour, and D. Vitali, \textsl{Quantum optomechanics of a multimode system coupled via a photothermal and a radiation pressure force}, Phys. Rev. A 86 (2012) 043803.
\bibitem{SN-P13} A.H. Safavi-Naeini, J. Chan, J.T. Hill, S. Gr\"oblacher, H. Miao, Y. Chen, M. Aspelmeyer, and M.O. Painter,  \textsl{Laser noise in cavity-optomechanical cooling and thermometry}, New J. Phys. \textbf{15} (2013) 035007 .

\bibitem{Chen13} Y. Chen, \textsl{Macroscopic quantum mechanics: theory and experimental concepts of optomechanics}, J. Phys. B: At. Mol. Opt. Phys. \textbf{46} (2013) 104001 .
\bibitem{asp14} M. Aspelmeyer, T.J. Kippenberg, and F. Marquardt, \textsl{Cavity optomechanics}, Rev. Mod. Phys. \textbf{86} (2014) 1391--1452 .
\bibitem{GroTru15} S. Gr\"oblacher, A. Trubarov, N. Prigge, G.D. Cole, M. Aspelmeyer, and J. Eisert,
    \textsl{Observation  of non-Markovian micro-mechanical Brownian motion}, Nat. Commun. 6  (2015) 7606.
\bibitem{BarV15} A. Barchielli and B. Vacchini, \textsl{Quantum Langevin equations for optomechanical systems}, New J.\ Phys.\ \textbf{17} (2015) 083004.
\bibitem{BM16} W.P. Bowen and G.J. Milburn, \textit{Quantum Optomechanics} (CRC, Taylor \& Francis, 2016).
\bibitem{Vit+18} P. Piergentili, L. Catalini, M. Bawaj, S. Zippilli, N. Malossi, R. Natali, D. Vitali, and G. Di Giuseppe, \textsl{Two-membrane cavity optomechanics}, New J.\ Phys. \textbf{20} (2018) 083024.
\bibitem{ZKRDV18}
S. Zippilli, N. Kralj, M. Rossi, G. Di Giuseppe, and D. Vitali, \textsl{Cavity optomechanics with feedback-controlled in-loop light}, Phys. Rev. A 98 (2018) 023828 .
\bibitem{Bar16} A. Barchielli, \textsl{Quantum stochastic equations for an opto-mechanical oscillator with radiation pressure interaction and
    non-Markovian effects}, Rep. Math. Phys. \textbf{77} (2016) 315--333.
\bibitem{HR06} S. Haroche and J.-M. Raimond, \textit{Exploring the Quantum. Atoms, Cavities and Photons} (Oxford University Press, Oxford, 2006).
\bibitem{WisM10} H.M. Wiseman and  G.J.  Milburn, \textit{Quantum Measurement and Control} (Cambridge University Press, Cambridge  2010).
\bibitem{M79} L. Mandel, \textsl{Sub-Poissonian photon statistics in resonance fluorescence}, Opt. Lett. \textbf{4} (1979) 205-207.
\bibitem{Car99} H.J. Carmichael, \textit{Statistical Methods in Quantum
    Optics}, Vol 1 (Springer, Berlin, 1999).
\bibitem{Car08} H.J. Carmichael, \textit{Statistical Methods in Quantum
    Optics}, Vol 2 (Springer, Berlin, 2008).
\bibitem{HudP84} R.L. Hudson and K.R. Parthasarathy, \textsl{Quantum It\^o's formula and  stochastic evolutions}, Commun. Math. Phys. \textbf{93} (1984) 301--323.
\bibitem{Parthas92} K.R. Parthasarathy, \textit{An Introduction to Quantum Stochastic  Calculus} (Birkh\"{a}user, Basel, 1992).

\bibitem{Bar06} A. Barchielli, \textsl{Continual Measurements in Quantum Mechanics and  Quantum Stochastic Calculus}, in \textit{Open Quantum Systems III},  S. Attal, A. Joye, and C.-A. Pillet eds.,
    Lect.\ Notes Math.\ \textbf{1882} (Springer, Berlin, 2006) pp. 207--291.
\bibitem{BarP02} A. Barchielli and N. Pero, \textsl{A quantum
    stochastic approach to the spectrum of a two-level atom}, J.\ Opt.\ B:
    Quantum Semiclass.\ Opt.\ \textbf{4} (2002) 272--282.
\bibitem{GJN12} J.E. Gough, M.R. James, and H.I. Nurdin,  \textsl{Single photon quantum filtering using non-Markovian embeddings}, Phil. Trans.
    R. Soc. A \textbf{370} (2012) 5408--5421.
\bibitem{BarGQP13} A. Barchielli and M. Gregoratti, \textsl{Entanglement protection and generation under continuous monitoring}, in L. Accardi, F. Fagnola,  \textit{Quantum  Probability and Related Topics}, QP-PQ: Quantum Probability and White
    Noise Analysis, Vol.\ 29, (World Scientific, Singapore, 2013) pp.\  17--42.
\bibitem{JG15} J.E. Gough, \textsl{Scattering processes in quantum optics}, Phys. Rev. A, \textbf{91} (2015) 013802.
\bibitem{ZGVit15} S. Zippilli, G. Di Giuseppe, and D. Vitali, \textsl{Entanglement and
squeezing of continuous-wave stationary light}, New J. Phys. \textbf{17} (2015) 043025.
\bibitem{ASth} A. Santamato, \textsl{A quantum theory of photodetection and other optical devices}, master thesis, University of Milan (2010). DOI 10.13140/RG.2.2.36655.48801.
\bibitem{ZolG97}  P. Zoller and C.W. Gardiner,   \textsl{Quantum noise in quantum optics: the
    stochastic Schr\"odinger equation}. In S. Reynaud, E. Giacobino \& J. Zinn-Justin eds., \textit{Fluctuations quantiques, (Les Houches 1995)} (North-Holland, Amsterdam, 1997) pp. 79--136.
\bibitem{GarC85} C.W. Gardiner and M.J. Collet, \textsl{Input and output in damped quantum systems: Quantum stochastic differential equations and the master equation}, Phys. Rev. A \textbf{31} (1985) 3761--3774.
\bibitem{Bar86} A. Barchielli, \textsl{Measurement theory and stochastic differential equations in quantum mechanics}, Phys.\ Rev.\ A \textbf{34} (1986) 1642--1649.

\bibitem{GarZ00} C.W. Gardiner and P. Zoller, \textit{Quantum Noise}, Springer Series in Synergetics, Vol.\ 56 (Springer, Berlin, 2000).
\bibitem{BarG08b} A. Barchielli and M. Gregoratti, \textsl{Quantum continual measurements: the   spectrum of the output}, in Quantum
    Probability and Related Topics, J.C.\ Garc\'{\i}a, R.\ Quezada, and S.\ B.\ Sontz eds., Quantum Probability Series QP-PQ Vol.\ 23, (World
    Scientific, Singapore, 2008) pp.\ 63--76.
\bibitem{BarG12} A. Barchielli and M. Gregoratti, \textsl{Quantum measurements in
    continuous time, non-Markovian evolutions and feedback}, Phil. Trans.
    R. Soc. A \textbf{370} (2012) 5364--5385.
\bibitem{BarG13} A. Barchielli and M. Gregoratti, \textsl{Quantum continuous measurements: The stochastic Schr\"o\-dinger equations and the spectrum of the output},  Quantum Measurements and Quantum Metrology \textbf{1} (2013) 34--56.
\bibitem{HKG19} D.B. Horoshko, M.I. Kolobov, F. Gumpert, I. Shand, F. K\"onig, M.V. Chekhova, \textsl{Nonlinear Mach-Zehnder interferometer with ultrabroadband squeezed light}, J. Mod. Opt. \textbf{67} (2020) 41-48.
\bibitem{Yur85} B. Yurke, \textsl{Wideband photon counting and homodyne detection} Phys. Rev. A \textbf{32} (1985) 311--323.
\bibitem{Leo10} U. Leonhardt, \textit{Essential Quantum Optics. From Quantum Measurements to
Black Holes} (Cambridge University Press, Cambridge, 2010)

\bibitem{WalM94} D.F. Walls and G.J. Milburn, \textit{Quantum Optics} (Springer, Berlin, 1994).
\bibitem{Greg01}
M. Gregoratti, \textsl{The Hamiltonian operator associated to some quantum stochastic
evolutions}, Commun. Math. Phys. \textbf{222} (2001) 181--200; \textsl{Erratum}, Commun. Math. Phys. \textbf{264} (2006) 563--564.
\bibitem{FagW03} F. Fagnola and S.J. Wills, \textsl{Solving quantum stochastic differential equations with unbounded coefficients}, J. Funct. Anal. \textbf{198} (2003) 279--310.
\bibitem{RS80} M. Reed, B. Simon, \textit{Methods of Modern Mathematical Physics II: Functional Analysis} (Academic Press, New York, 1980).
\bibitem{Lin76} G. Lindblad, \textsl{On the generators of quantum dynamical semigroups}, Commun. Math. Phys. \textbf{48} (1976) 119--130.
\bibitem{Lin76b} G. Lindblad, \textsl{Brownian motion of a quantum harmonic oscillator}, Rep. Math Phys. \textbf{10} (1976) 393--407.
\bibitem{BarCS11} R. Castro Santis and A. Barchielli, \textsl{Quantum
    stochastic differential  equations and continuous measurements:
    unbounded coefficients}, Rep. Math. Phys. \textbf{67} (2011) 229--254.
\bibitem{GarPC87} C.W. Gardiner, A.S. Parkins and M.J. Collet, \textsl{Input and output in damped quantum systems. II. Methods
in non-white-noise situations and application to inhibition of
atomic phase decays}, J. Opt. Soc. Am. B \textbf{4} (1987) 1683--1699.

\bibitem{Meyer93} P.A. Meyer, \textit{Quantum probability for probabilists}. Lecture Notes in
    Mathematics \textbf{1538} (Springer-Verlag, Berlin, 1993).
\bibitem{Brun02} T.A. Brun, \textsl{A simple model of quantum trajectories} American Journal of Physics \textbf{70} (2002) 719.
\bibitem{AttP06} S. Attal and Y. Pautrat, \textsl{ From Repeated to Continuous Quantum Interactions}, Ann. Henri Poincaré \textbf{7} (2006) 59--104.
\bibitem{Greg15} M. Gregoratti, \textsl{The Hamiltonian generating quantum stochastic evolutions in the limit from repeated to continuous interactions},  Open Systems \& Information Dynamics, \textbf{22}  (2015) 1550022.
\bibitem{Cicc17} F. Ciccarello, \textsl{Collision models in quantum optics}, Quantum Meas. Quantum Metrol. \textbf{4} (2017) 53--63.
\bibitem{MEsp17}  P. Strasberg, G. Schaller, T. Brandes, and  M. Esposito, \textsl{Quantum and information thermodynamics: a unifying framework based on repeated interactions}, Phys. Rev. X \textbf{7} (2017) 021003.
\bibitem{Giov18} S. Cusumano, V. Cavina, M. Keck, A. De Pasquale, and V. Giovannetti, \textit{Entropy production and asymptotic factorization via thermalization: A collisional model approach}, Phys. Rev. A \textbf{98} (2018) 032119.
\bibitem{GiovSciarrMata19} A. Cuevas, A. Geraldi, C. Liorni, L.D. Bonavena, A. De Pasquale, F. Sciarrino, V. Giovannetti, and P. Mataloni, \textsl{All-optical implementation of collision-based evolutions of open quantum systems}, Scientific Reports (2019) 9:3205.
\bibitem{Bel12} V.P. Belavkin, \textsl{Quantum demolition filtering and optimal control of unstable systems}, Phil. Trans.
    R. Soc. A \textbf{370} (2012) 5396--5407.
\bibitem{GJ09} J. Gough and M.R. James, \textsl{The series product and its application to quantum
feedforward and feedback networks}, IEEE Trans. Aut. Control \textbf{ 54} (2009) 2530--2544.

\end{thebibliography}
\end{document}